\documentclass[12pt]{article}
\usepackage{amsmath,amssymb,epsfig,amsfonts}
\usepackage{graphicx,subfigure}
\usepackage[usenames, dvipsnames]{color}
\usepackage[backref]{hyperref}
\usepackage{cite}
\usepackage{verbatim} %for comments

\makeatletter

\newcommand{\be}{\begin{equation}}
\newcommand{\ee}{\end{equation}}
\newcommand{\ba}{\begin{aligned}}
\newcommand{\ea}{\end{aligned}}

\newcommand{\C}{\mathbb{C}}
\renewcommand{\P}{\mathbb{P}}

\newcommand{\cL}{\mathcal{L}}

\newcommand{\cN}{\mathcal{N}}

\newcommand{\bea}{\begin{eqnarray}}
\newcommand{\eea}{\end{eqnarray}}

\newcommand{\R}{{\mathbb R}}
\newcommand{\Z}{{\mathbb Z}}

\def\unit{{1\kern-.65ex {\rm l}}}
\def\1{{1\kern-.65ex {\rm l}}}

\newcount\hour \newcount\minute
\hour=\time \divide \hour by 60
\minute=\time
\def\now{%
\ifnum \hour<13
  \ifnum \hour=0 \advance \hour by 12 \number\hour:\else \number\hour:\fi%
     \ifnum \minute<10 0\fi%
     \number\minute%
\ A.M.%
\else \advance \hour by -12 \number\hour:%
  \ifnum \minute<10 0\fi%
  \number\minute%
  \ P.M.%
\fi%
}

\makeatother

\begin{document}

% format
\baselineskip=18pt  % a la harvmac
\numberwithin{equation}{section}  % make eq labels (sec.num)
\allowdisplaybreaks  % allow page breaks in displayed eqs

%%%%%%%%%%%%%%%%%%%%%%%%%%%%%%%%%%%%%%%%%%%
%%%        TITLE BEGINS HERE
%%%%%%%%%%%%%%%%%%%%%%%%%%%%%%%%%%%%%%%%%%%

%% ========== title (note version) begins here ==========
%
%\vspace*{-1cm}
%\begin{center}
% {\Large\bf Title of the Document}
%\end{center}
%\vspace*{-.5cm}
%
%% ========== title (note version) ends here ==========

%% ========== title (paper version, a la harvmac) begins here ==========

\thispagestyle{empty}

% Report number
\vspace*{-2cm} 
\begin{flushright}
KCL-MTH-13-02\\
KIAS P-13017 
\end{flushright}

% title, authors, affiliation
\vspace*{0.8cm} 
\begin{center}
 {\LARGE Phases, Flops and F-theory:\\
  \smallskip
 $SU(5)$ Gauge Theories}\\

 \vspace*{1.5cm}
{Hirotaka Hayashi$^1$, Craig Lawrie$^2$ and Sakura Sch\"afer-Nameki$^2$}\\

 \vspace*{1.0cm} 
$^1$  {\it School of Physics, Korea Institute for Advanced Study,\\
 Seoul 130-722, Korea}\\
{\tt hayashi kias.re.kr}\\
\vspace*{0.5cm}
$^2$ {\it Department of Mathematics, King's College, London \\
  The Strand, London WC2R 2LS, England }\\
  {\tt {gmail:$\,$ craig.lawrie1729, sakura.schafer.nameki}}

\vspace*{0.8cm}
\end{center}
\vspace*{.5cm}

% abstract
\noindent
We consider  F-theory and M-theory compactifications on singular Calabi-Yau fourfolds with an $SU(5)$ singularity.
On the M-theory side this realizes three-dimensional $\mathcal{N}=2$ supersymmetric gauge theories with matter, and compactification on a resolution of the fourfold corresponds to passing to the Coulomb branch of the gauge theory. 
The classical phase structure of these theories has a simple characterization in terms of subwedges of the fundamental Weyl chamber of the gauge group. This phase structure has a counterpart in the network of small resolutions of the Calabi-Yau fourfold. We determine the geometric realization of each phase, which crucially depends on the fiber structure in codimension 2 and 3, including the network structure, which is realized in terms of flop transitions. This results in a set of small resolutions, which do not have a standard algebraic or toric realization, but are obtained by flops along codimension 2 (matter) loci. 

\newpage
%%%%%%%%%%%%%%%%%%%%%%%%%%%%%%%%%%%%%%%%%%%
%%%           TITLE ENDS HERE
%%%%%%%%%%%%%%%%%%%%%%%%%%%%%%%%%%%%%%%%%%%

\tableofcontents
%\printindex

%%%%%%%%%%%%%%%%%%%%%%%%%%%%%%%%%%%%%%%%%%%
%%%        MAIN TEXT BEGINS HERE
%%%%%%%%%%%%%%%%%%%%%%%%%%%%%%%%%%%%%%%%%%%
\newpage

%%%%%%%%%%%%%%%%%%%%%%%%%%%%%%%%%%%%
%%%%%%%%%%%%%%%%%%%%%%%%%%%%%%%%%%%%

\section{Introduction}

F-theory provides an ideal setting for geometric engineering of  gauge theories \cite{Vafa:1996xn, Morrison:1996na, Morrison:1996pp}. The singularity structure of elliptically fibered Calabi-Yau manifolds translates into gauge theoretic data at low energies, which describe the effective theory of 7-branes wrapping internal cycles of the Calabi-Yau space. 
The Kodaira type of the singular elliptic fibers in codimension 1 translates into the gauge group of the low energy {effective} theory. 
In addition to the gauge degrees of freedom, matter and Yukawa couplings can be realized in terms of singularities that occur in higher codimension in the base of the elliptic fibration. This has in particular played a key role in the recent surge in construction of grand unified theories, mainly based on the gauge group $SU(5)$, in four dimensions by compactifying F-theory on Calabi-Yau fourfolds 
(for a nice review see \cite{Weigand:2010wm}). The fiber structure that arises in this context has been determined by explicit resolution of the singularities utilizing various approaches -- resolution in local patches, or in terms of global sections, or toric resolutions  \cite{Esole:2011sm, MS, Krause:2011xj, Grimm:2011fx, Lawrie:2012gg, Braun:2013cb} -- with the main motivation to study the structure of Yukawa couplings and construction of $G_4$-fluxes. As has been noted  
in some of these works, the resolution is not unique, and there is a network of so-called small resolutions. The structure of these networks is tied to the higher-codimension singularities, that can occur in Calabi-Yau fourfolds. For instance, the fiber in codimension 1, which is generically an $I_5$ Kodaira fiber, and which is independent of the small resolution, 
can further degenerate along the higher codimension singular loci. The small resolution determines how the fiber splits along these loci, which on the other hand determines the phase of the three-dimensional gauge theory \cite{Grimm:2011fx, Cvetic:2012xn}. 
In this paper we will present a complete picture of the types of small resolution networks that can arise, and match this with the gauge theoretic phase structure. 

The inspiration for the structure of small resolutions comes from the dual M-theory compactification, and the phase structure of the resulting three-dimensional gauge theories.
Indeed, the resolution of singular elliptic Calabi-Yau manifolds also plays an important role 
from the perspective of F-theory/M-theory duality \cite{Morrison:1996na, Morrison:1996pp}, in particular in {relation to} three-dimensional $\mathcal{N}=2$ supersymmetric gauge theories with matter multiplets introduced in \cite{deBoer:1997kr, Aharony:1997bx}. 
M-theory compactifications on Calabi-Yau fourfolds yield three-dimensional $\mathcal{N}=2$ supersymmetric gauge theories, where the codimension 1 singularity determines the gauge group, and the resolution of the singularity corresponds to going to a Coulomb branch of the three-dimensional theory. 
{This correspondence can also be useful in understanding various aspects of F-theory compactifications (see for instance \cite{Denef:2008wq} for a review). }

In the present context we will {determine} the Coulomb branch of three-dimensional $\mathcal{N}=2$ gauge theories in order to obtain information {on the types of resolutions of the  Calabi-Yau fourfolds}. Since the analysis of the classical Coulomb branch is generic and systematic, one obtains a complete picture of the resolution network. 

{
In order to make contact with $SU(5)$ F-theory compactifications, we will mainly focus on the Coulomb branch of $d=3$,  $\mathcal{N}=2$, $SU(5)$ gauge theories with matter fields in the fundamental ${\bf 5}$ and the anti-symmetric ${\bf 10}$ representation. The inclusion of the anti-symmetric representation enriches the structure of the Coulomb branch.
In the present context, we are interested in the different resolutions of the geometric singularities. Therefore we will focus on the classical Coulomb branch\footnote{The effect of $G_4$ flux on the phase structure of these theories was studied in \cite{Intriligator:2012ue}.}, which reveals the complete set of phases of the three-dimensional  theory. We characterize the phases in terms of subwedges in the fundamental Weyl chamber. 
Geometrically, we find that 
two phases that share a codimension 1 hyperplane in the classical Coulomb moduli space will be connected by flop transitions. 
The Coulomb branch of the three-dimensional gauge theories predict phases, which had not previously been constructed geometrically. We construct the resolutions corresponding to these new phases by explicitly performing the flop transitions.

Concretely, starting with this network of phases derived from a purely gauge theoretic point of view, we find geometric realizations of these using first toric resolutions as in \cite{Krause:2011xj, Grimm:2011fx, Grimm:2009yu}, as well as algebraic resolutions as in \cite{MS, Lawrie:2012gg}. } Both toric and algebraic methods cover only parts of the network of phases, however the latter can be used to do explicit flop transitions, which then yields the remaining {gauge theory} phases. {It is noteworthy, that these new resolutions do not have a toric realization, as they arise from flops along matter curves in codimension 2, which are not intersections of two exceptional divisors. There is a beautiful connection between gauge theory phases, representation theory and networks of flops which will be discussed in general in \cite{HLS}. 
}

The organization of the paper is {as} follows. In section 2, we briefly review the Coulomb branch of three-dimensional $\mathcal{N}=2$ gauge theories and determine these in the case of $SU(5)$ gauge theories with ${\bf 5}$ and ${\bf 10}$ matter fields. We also describe how to identify the Coulomb branch with the geometric resolution of singularities in Calabi-Yau fourfolds. Sections 3 and 4 discuss the geometric resolutions of the $SU(5)$ singularity in codimension 1, 2 and 3 using toric and algebraic methods, respectively. Neither of these generate the full phase structure that was observed in section 2. 
In section 5, we explicitly construct new {resolutions of the geometry, as} predicted from the three-dimensional gauge theories starting with the algebraic resolutions,  and then performing flop transitions along components of the fiber in codimension 2. Some of the technical details in section 3 and 4 are relegated to the appendices.

%\newpage
%%%%%%%%%%%%%%%%%%%%%%%%%%%%%%%%%%%%
%%%%%%%%%%%%%%%%%%%%%%%%%%%%%%%%%%%%

\section{Coulomb Phases of $d=3$ $\cN=2$ Gauge Theories}
\label{sec:phase}

%%%%%%%%%%%%%%%%%%%%%%%%%%%%%%%%%%%%
\subsection{General structure}

We first review the general structure of the Coulomb branch of three-dimensional $\cN=2$ gauge theories with 
$N_f$ chiral multiplets $Q_f$ in a representation ${\bf R}_f$ of a gauge group $G$ \cite{deBoer:1997kr, Aharony:1997bx}. We assume that there are {neither} classical real mass terms nor classical complex mass terms for the chiral multiplets in order to make contact with the subsequent {purely geometric} analysis. We also set the classical Chern-Simons term to zero. The three-dimensional $\cN=2$ vector multiplet $V$ has a real scalar denoted by $\phi$ in the adjoint representation of $G$, which plays an important role.  

The gauge theory has a Coulomb branch, where the real scalar $\phi$ picks up a vacuum expectation value (vev) $\langle\phi \rangle$ in the Cartan subalgebra of $G$, and the gauge group generically breaks {to} $U(1)^{r}$, where $r={\rm rank}(G)$. The Coulomb branch is then described as the Weyl chamber $\R^{r}/W$, where $W$ is the Weyl group of $G$.  Without loss of generality, we consider the fundamental Weyl chamber in the following, which is characterized by
\be
\alpha_i \cdot \phi > 0, \qquad i=1, \cdots, r \,, 
\label{weyl}
\ee
where $\alpha_i$ denotes a simple root of $G$. It is useful to write the scalars in the Cartan subalgebra in terms of components $\phi^i$ that are written in the basis of fundamental weights (i.e. the duals to the simple coroots). Then, the product in \eqref{weyl} is defined by the quadratic form matrix. At the boundary of the Coulomb branch, some of the abelian gauge symmetries enhance to non-abelian ones at the classical level.

In the bulk of the Coulomb branch, one may complexify the real scalar $\phi^i$ by a scalar $\gamma^{i},\quad i= 1, \cdots, r$ dual to the photons associated to $U(1)^{r}$ gauge fields. The fields $\gamma^{i}$ live on an $r$-dimensional torus because of charge quantization. The complex scalar $\Phi^i = \phi^i + i\gamma^i$ is  then a scalar component of a chiral superfield. Due to the identification along the torus, the single valued superfields are $e^{\frac{{\bf r}_i\cdot\Phi}{g^2}}$ where $g$ is the gauge coupling, which also characterizes the size of the torus. 
%One can also construct a current $J^{(j)}_{\mu} = \epsilon_{\mu\nu\rho}(F^{\mu\nu})^{(j)}$ which generates $U(1)_{J}^{r}$ global symmetries. Here $F_{\mu\nu}^{(i)}$ denotes the field strength for $U(1)^{r}$ gauge symmetries and the Greek letters run from $0$ to $2$. The global symmetries induce shifts of $\gamma^i$. 

In the presence of chiral multiplets $Q_f$, there is a substructure in the fundamental Weyl chamber. The classical Lagrangian has terms 
\be
\cL \supset \sum_f |\phi Q_f|^2,
\ee
where $\langle\phi\rangle$ behaves as a real mass term for the chiral multiplet $Q_f$. Suppose $Q_f$ carries a weight ${\bf w}_f$ in a representation ${\bf R}_f$, then we have mass terms     
\be
\cL \supset \sum_f |\phi\cdot{\bf w}_f|^2 |Q_f|^2.
\ee
Therefore, there appear additional massless matter fields at least classically along the boundary where $\phi\cdot{\bf w}_f = 0$ inside the fundamental Weyl chamber \eqref{weyl}. Hereby, the fundamental Weyl chamber is further divided into subwedges under the inclusion of chiral multiplets in non-trivial representations. The boundary is a real codimension 1 locus in the $2r$-dimensional space. At the boundary $\phi\cdot{\bf w}_f = 0$, one can also turn on a vev $\langle Q_f \rangle \neq 0$ which corresponds to a Higgs branch, which intersects with the Coulomb branch at the boundary. This is the basic picture of the classical Coulomb branch of three-dimensional $\mathcal{N}=2$ gauge theories.

Quantum corrections may alter the structure of the classical Coulomb branch. Along the boundary $\phi\cdot{\bf w}_f = 0$, the low energy theory is governed by a particular $U(1)$ symmetry with massless $N_f$ flavors. It is argued that perturbative quantum corrections make the radius of the torus vanish along the locus, and the Coulomb branch splits into two regions where different variables become valid on either side of the boundary \cite{deBoer:1997kr, Aharony:1997bx}. Hence,  quantum mechanically, we have a complex codimension 1 boundary.  Indeed, the number of fermionic zero modes in an monopole background\footnote{In the Coulomb branch of non-Abelian gauge theories, there can be monopoles associated to $\pi_2(G/U(1)^r) = \mathbb{Z}^r$. They are essentially the same as the four-dimensional monopoles in \cite{Weinberg:1979zt}.} can change across the boundary. The monopoles can generate non-perturbative superpotentials in some of the subwedges, where we have an appropriate number of the fermionic zero modes. This suggests that the superpotential generated by the monopole can jump along the complex codimension 1 boundary. Also, some of the subwedges in the Coulomb branch will be lifted by the {superpotential} \cite{Affleck:1982as, deBoer:1997kr, Aharony:1997bx}. Furthermore, if the matter content is not vector-like, the Chern-Simons term is generated by a one-loop effect \cite{Aharony:1997bx}, which also lifts the Coulomb branch \cite{Deser:1981wh, Deser:1982vy}. 

In the geometric analysis, which will be discussed later, we will simply consider the resolutions of singular elliptically fibered Calabi-Yau fourfolds in the M-theory compactification. The non-perturbative superpotentials are generated by M5-brane instantons \cite{Witten:1996bn} and the Chern-Simons term is generated by the background $G_4$-flux, which can also induce non vector-like spectra. Without the effects of M5-brane instantons or $G_4$-flux, the Coulomb branch may not be lifted\footnote{The connection between the degeneration of Calabi-Yau fourfolds and the non-perturbative superpotential in some models was discussed in \cite{Katz:1996th, Diaconescu:1998ua}.}. In particular, we will see that the K\"ahler cone of the resolved Calabi-Yau fourfold is related to the classical Coulomb branch described by the Cartan scalars $\phi^i$. Hence, the classical analysis of the Coulomb branch is enough to discuss the resolution of singularities in elliptically fibered Calabi-Yau fourfolds.

%%%%%%%%%%%%%%%%%%%%%%%%%%%%%%%%%%%%
\subsection{$SU(5)$ Gauge Theories with Matter Representations}
\label{sec:Weights}

Motivated by the recent studies of $SU(5)$ singularities in Calabi-Yau fourfolds (for instance in the context of GUT model building), we now turn to this specific example and determine its phase structure of the classical Coulomb branch. More precisely, consider an $SU(5)$ gauge theory with $N_{{\bf 5}}$ chiral multiplets in the fundamental representation and anti-fundamental representation and $N_{{\bf 10}}$ chiral multiplets in the anti-symmetric representation and its complex conjugate representation. 

\begin{figure}[tb]
\begin{center}
\includegraphics[width=80mm]{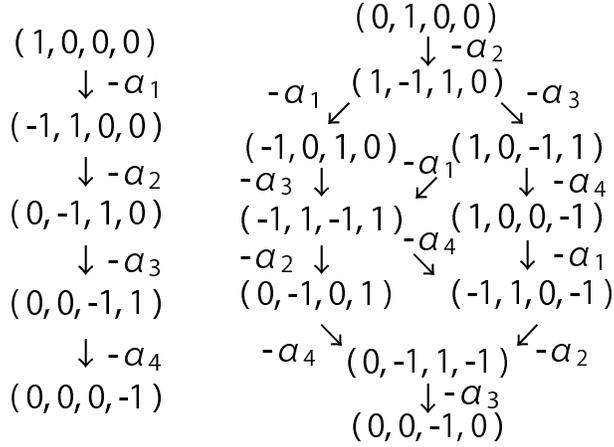}
\end{center}
\caption{The left/right figure shows the weights in terms of Dynkin labels of the  {\bf 5}/{\bf 10} representation.}
\label{fig:weight}
\end{figure}

To setup some notation for the phase structure of the theory, we summarize the weights of the 
${\bf 5}$ and ${\bf 10}$ representations in figure \ref{fig:weight}. 
%%%
%%%%
Furthermore, denote the simple roots of the $SU(5)$ Lie algebra using Dynkin labels by
\be
\alpha_1 = (2,-1,0,0),\;\; \alpha_2=(-1,2,-1,0),\;\; \alpha_3=(0,-1,2,-1),\;\;\alpha_4=(0,0,-1,2) \,.
\ee
It will be useful to give each of the weight in figure \ref{fig:weight} labels
\be
{\bf w}^{{\bf 5}}_1 = (1,0,0,0),\;\; {\bf w}^{{\bf 5}}_2=(-1,1,0,0),\;\; {\bf w}^{{\bf 5}}_3=(0,-1,1,0),\;\;{\bf w}^{{\bf 5}}_4=(0,0,-1,1),\;\;{\bf w}^{{\bf 5}}_5=(0,0,0,-1) \,,
\ee
for the {\bf 5} representation and 
\bea
{\bf w}^{{\bf 10}}_1&=&(0,1,0,0),\;\;{\bf w}^{{\bf 10}}_2=(1,-1,1,0),\;\;{\bf w}^{{\bf 10}}_3=(1,0,-1,1),\;\;{\bf w}^{{\bf 10}}_4=(1,0,0,-1),\\
{\bf w}^{{\bf 10}}_5&=&(-1,0,1,0),\;\; {\bf w}^{{\bf 10}}_6=(-1,1,-1,1),\;\;{\bf w}^{{\bf 10}}_7=(-1,1,0,-1),\\
{\bf w}^{{\bf 10}}_8&=&(0,-1,0,1),\;\;
{\bf w}^{{\bf 10}}_9=(0,-1,1,-1),\;\;{\bf w}^{{\bf 10}}_{10}=(0,0,-1,0) \,,
\eea
for the {\bf 10} representation.

%%%%%%%%%%%%%%%%%%%%%%
\begin{table}[t]
\begin{center}
\begin{tabular}{|c|c|c|c|c|c|}
\hline
Phase $\setminus$ Weight & ${\bf w}^{{\bf 5}}_1$ &  ${\bf w}^{{\bf 5}}_2$ &  ${\bf w}^{{\bf 5}}_3$ &  ${\bf w}^{{\bf 5}}_4$ &  ${\bf w}^{{\bf 5}}_5$  \\
\hline
I & + & + & + & + & -  \\
\hline
II & + & + & + & - & -  \\
\hline
III & + & + & - & - & -  \\
\hline
IV & + & - & - & - & -  \\
\hline
\end{tabular}
\end{center}
\caption{All the phases of the $SU(5)$ gauge theories with chiral multiplets in the fundamental representation and the anti-fundamental representation.}
\label{tb:5}
\end{table}
%%%%%%%%%%%%%%%%%%%%%%%%%%%%%%%%%%
%%%%%%%%%%%%%%%%%%%%%%%%%%%%%%%%%%

\begin{table}[t]
\begin{center}
\begin{tabular}{|c|c|c|c|c|c|c|c|c|c|c|}
\hline
Phase $\setminus$ Weight & ${\bf w}^{{\bf 10}}_1$ &  ${\bf w}^{{\bf 10}}_2$ &  ${\bf w}^{{\bf 10}}_3$ &  ${\bf w}^{{\bf 10}}_4$ &  ${\bf w}^{{\bf 10}}_5$ &  ${\bf w}^{{\bf 10}}_6$ &  ${\bf w}^{{\bf 10}}_7$ &  ${\bf w}^{{\bf 10}}_8$ &  ${\bf w}^{{\bf 10}}_9$ &  ${\bf w}^{{\bf 10}}_{10}$ \\
\hline
I' & + & + & + & + & + & + & + & - & - & - \\
\hline
II' & + & + & + & + & + & + & - & - & - & - \\
\hline
III' & + & + & + & + & + & - & - & - & - & - \\
\hline
IV' & + & + & + & + & - & - & - & - & - & - \\
\hline
V' & + & + & + & - & + & + & - & + & - & - \\
\hline
VI' & + & + & + & - & + & + & - & - & - & - \\
\hline
VII' & + & + & + & - & + & - & - & - & - & - \\
\hline
VIII' & + & + & - & - & + & - & - & - & - & - \\
\hline
\end{tabular}
\end{center}
\caption{All the phases of the $SU(5)$ gauge theories with chiral multiplets in the anti-symmetric representation and its complex conjugate representation.}
\label{tb:10}
\end{table}
%%%%%%%%%%%%%%%%%%%%%%%%%%%%%%%%%%%%%%%%%%%%%%
\begin{table}[t]
\begin{center}
\begin{tabular}{|c|c|c|c|c|c|c|c|c|c|c|c|c|}
\hline
Repr. $\setminus$ Phase & 1 & 2 & 3 & 4 & 5 &  6 & 7 & 8 & 9 & 10 & 11 & 12 \\
\hline
{\bf 5} & III & III & II & III & III & IV & I & II & II & III & II & II\\
\hline
{\bf 10} & I' & II' & III' & III' & IV' & IV' & V' & V' &  VI'& VI' & VII' & VIII' \\
\hline
\end{tabular}
\end{center}
\caption{All the phases of the $SU(5)$ gauge theories with chiral multiplets in the fundamental representation and the anti-symmetric representation. Each column represents a possible phase combination.}
\label{tb:phase}
\end{table}
%%%%%%%%%%%%%%%%%%%%%%%%%%%%%%%

In order to classify all the phases of the $SU(5)$ gauge theory, one determines a non-empty region in the fundamental Weyl chamber, which satisfies either $\phi\cdot{\bf w}_f > 0$ or $\phi\cdot{\bf w}_f < 0$ for each weight of ${\bf 5}$ and ${\bf 10}$ representations. Note that not all the combinations of $\phi\cdot{\bf w}_f > 0$ or $\phi\cdot{\bf w}_f < 0$ are allowed. For example, there is no region in the fundamental Weyl chamber which satisfies $\phi \cdot {\bf w}_f > 0$ for all the weights of the fundamental representation of $SU(5)$. For simplicity, we use the notation
\be\ba
{\bf w}_f > 0 &\qquad \leftrightarrow\quad \phi\cdot{\bf w}_f > 0\cr
{\bf w}_f < 0 &\qquad \leftrightarrow\quad  \phi\cdot{\bf w}_f < 0 \,.
\ea\ee
We first choose the Coulomb branch to lie in the fundamental Weyl chamber \eqref{weyl},
\be
\alpha_1 > 0,\;\; \alpha_2 >0,\;\; \alpha_3 >0,\;\; \alpha_4 > 0.
\ee
Inside this fundamental Weyl chamber, there is a subwedge structure due to the presence of the chiral multiplets in the fundamental representation, the anti-symmetric representation and their complex conjugate representations. First, let us see the phases from the {\bf 5} representation. There are four phases altogether as in Table \ref{tb:5}.
On the other hand, we have eight phases from the {\bf 10} representation as in Table \ref{tb:10}.

However, not all combinations of the phases from the {\bf 5} representation and  {\bf 10} representation are possible,
{as the intersection of the regions characterizing the two phases can be empty.} In fact, we find twelve phases in total, {which are compatible,  and} are summarized in Table \ref{tb:phase}.
Each of these twelve phases corresponds to a subwedge in the fundamental Weyl chamber. One can also consider a cone in the weight space which is dual to the subwedge of the fundamental Weyl chamber. The cone is spanned by the weights or roots of the $SU(5)$. All the weights ${\bf w}$ inside the cone should satisfy ${\bf w} > 0$. The generators of the cones for all the phases are summarized in Table \ref{tb:generators}. 
All the cones are simplicial since the weight space has four real dimensions and the number of  generators for all the phases is also four.

%%%%%%%%%%%%%%%%%%%%%%%%%%%%%%%

\begin{table}[t]
\begin{center}
\begin{tabular}{|c|c|} 
\hline
Phase & Generators\\
\hline
1 & (2,-1,0,0),\;\; (0,-1,2,-1),\;\; (0,0,-1,2),\;\; (-1,1,0,-1),\\
\hline
2 & (0,-1,2,-1),\;\; (1,0,0,-1),\;\; (-1,1,-1,1),\;\; (1,-1,0,1),\\
\hline
3 & (-1,2,-1,0),\;\; (0,0,-1,2),\;\; (1,0, 0, -1),\;\; (0,-1, 1, 0),\\
\hline 
4 & (0,0,-1,2),\;\; (-1,0,1,0),\;\; (1,-1,1,-1),\;\;(0,1,-1,0),\\
\hline
5 & (0,-1,2,-1),\;\; (0,0,-1,2),\;\; (-1,1,0,0),\;\;(1,0,-1,0),\\
\hline
6 & (-1,2,-1,0),\;\; (0,-1,2,-1),\;\;(0,0,-1,2),\;\; (1,-1,0,0),\\
\hline 
7 & (2,-1,0,0),\;\;(-1,2,-1,0),\;\; (0,-1,2,-1),\;\; (0,0,-1,1),\\
\hline
8 & (2,-1,0,0), \;\; (-1,2,-1,0),\;\;(0,0,1,-1),\;\; (0,-1,0,1),\\
\hline
9 & (2,-1,0,0), \;\; (-1,1,-1,1),\;\;(0,-1,1,0),\;\;(0,1,0,-1),\\
\hline
10 & (2,-1,0,0),\;\; (0,-1,2,-1),\;\;(-1,0,0,1),\;\; (0,1,-1,0),\\
\hline 
11 & (-1,2,-1,0),\;\; (1,0,-1,1),\;\;(-1,0,0,1),\;\;(1,-1,1,-1),\\
\hline
12 & (2,-1,0,0),\;\; (-1,2,-1,0),\;\;(0,0,-1,2),\;\;(-1,0,1,-1),\\
\hline
\end{tabular}
\end{center}
\caption{Generators of the cone in the weight space for the 12 phases in table \ref{tb:phase}.}
\label{tb:generators}
\end{table}
%%%%%%%%%%%%%%%%%%%%%%%%%%%%%%%

\begin{figure}[tb]
\begin{center}
\includegraphics[width=80mm]{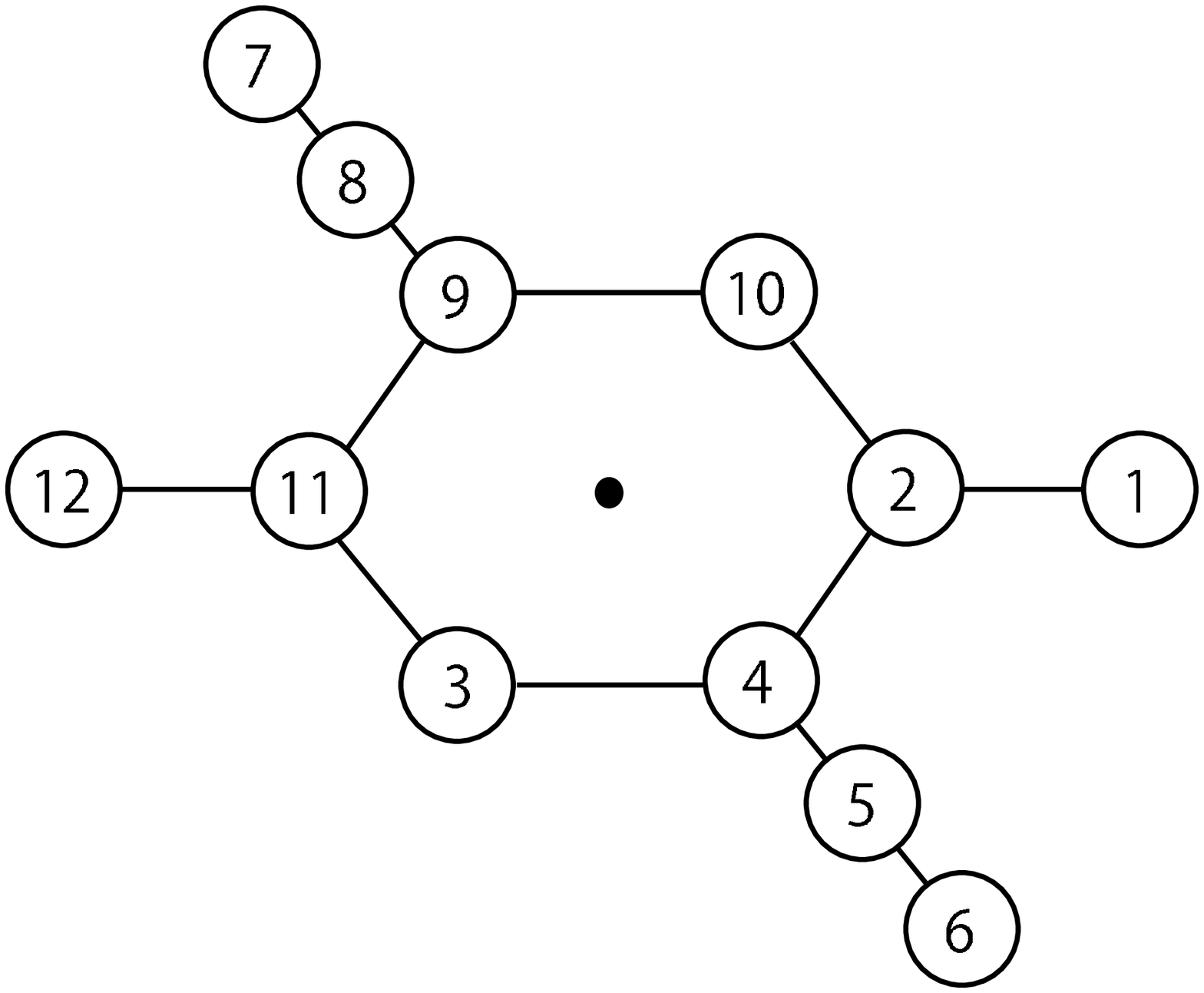}
\end{center}
\caption{The diagram showing the relation between the phases. Each circle represents a phase and the straight line between the circles show which phases share a common real codimension 1 wall. The central hexagon has a realization in terms of algebraic resolutions of the type $(ij)(kl)$ as explained in section \ref{sec:AlgRes}. The phases outside of the hexagon are realized in terms of flops in the algebraic resolutions in section \ref{sec:Flops}. There is a symmetry, which is reflection along the central dot, which amounts to a relabeling of  the Cartan generators.}
\label{fig:flop}
\end{figure}

 {Next, we consider the relations  among the 12 phases}. We will determine, which two phases share a real codimension 1 hyperplane. The wall of each phase is characterized by a weight ${\bf w}$, which satisfies  $\phi \cdot {\bf w} = 0$.  In order to see which two phases are adjacent, it is convenient to look at the generators of the cone in the weight space. If one phase has a generator ${\bf w}$, whose negative is a generator in another phase, then the two phases share a real codimension 1 wall, which is expressed by $\phi \cdot {\bf w} = 0$. For example,  phase 1 and  phase 2 are next to each other and the hyperplane shared by the two phases is $\phi \cdot (-1,1,0,-1) = 0$. This is because one of the generators, $(-1,1,0,-1)$, for phase 1 is exactly the negative of $(1,-1,0,1)$ in phase 2. In this way, one can relate all the phases in Table \ref{tb:phase} by checking the generators in the weight space in Table \ref{tb:generators}. The relations between each phase are summarized in Figure \ref{fig:flop}. 
%%%%
%%%%

%%%%%%%%%%%%%%%%%%%%%%%%%%%%%%%%%%%%

\subsection{Relation between {Gauge Theory Phases and Geometry}}

In order to make contact with the three-dimensional $\cN=2$ gauge theories with chiral multiplets in a representation ${\bf R}_f$ of a gauge group $G$, we consider an M-theory compactification on a Calabi-Yau fourfold $X_4$ which has a singularity of $G$ type over a complex surface $S$\footnote{Non-simply laced gauge groups can be realized by the monodromy associated with a loop in $S$, which corresponds to the outer automorphism of $ADE$ gauge groups \cite{Aspinwall:1996mw, Bershadsky:1996nh}.}. The matter can be encoded by the singularity enhancement loci on the complex surface \cite{Bershadsky:1996nh, Katz:1996xe}. The enhanced singularity type characterizes the representation of the matter fields localized along the loci. One can realize the Coulomb branch of the gauge theories by an M-theory compactification on such a Calabi-Yau fourfold $\tilde{X}_4$, where the singularities have been resolved.

{In the resolved geometry, there are} additional holomorphic curves $\Sigma \in H_2(\tilde{X}_4, \Z)$, which vanish in the singular limit. The matter fields, which become massive in the Coulomb branch, have a natural interpretation as M2-branes wrapping these holomorphic curves associated with the resolution of the singularity along the singularity enhancement loci in $S$. More precisely, 
 the matter fields arise from surfaces that have the structure of a curve (here denoted by $\Sigma$) fibered over a matter curve inside $S$.

 The roots of the original non-Abelian gauge symmetry $G$ can be also interpreted by M2-branes wrapping these holomorphic curves associated with the resolution of the singularity over $S$. Hence, the holomorphic curves may be labeled by weights of some representations of $G$ and denoted by $\Sigma_{{\bf w}}$. We will call the space of a collection of holomorphic curves which shrink in the singular limit the relative Mori cone $M(\tilde{X}_4 / X_4)$ \cite{Grimm:2011fx}. The dual cone of the relative Mori cone is called the relative K\"ahler cone $K(\tilde{X}_4/ X_4)$, which is defined as 
\be
K(\tilde{X}_4/ X_4) = \{D = \sum \phi^{i} D_{i} \; | \; D \cdot \Sigma_{\bf w} > 0 {\rm \;\;for\;\;all\;\;}\Sigma_{\bf w} \in M(\tilde{X}_4/X_4)\}\,,
\ee
where $D_i, i = 1, \cdots, {\rm rank}(G)$ are divisors in the resolved Calabi-Yau fourfold. These divisors {are given by} a holomorphic curve $\Sigma_{-\alpha_i}$ fibered over $S$, where $\Sigma_{-\alpha_i}$ is labeled by minus a simple root of $G$ and shrinks in the singular limit.  

The effective action from M-theory compactifications on Calabi-Yau fourfolds has been studied in \cite{Becker:1996gj, Gukov:1999ya, Dasgupta:1999ss, Haack:2001jz, Grimm:2010ks, Intriligator:2012ue}. The effective action has the three-dimensional $\cN=2$ vector multiplets whose bosonic components are 
\be
(\phi^{i}, A^i), \qquad i=1, \cdots, {\rm rank}(G).
\ee
The three-dimensional gauge field $A^i$ comes from the dimensional reduction of the M-theory three-form
\be
C_3 = A^i \wedge \omega_i, 
\ee
where $\omega_i$ is a Poincar\'e dual two-form to the divisor $D_i$. The real scalar $\phi^i$ can be obtained by the expansion of the normalized K\"ahler form $\tilde{J} = J \cdot \mathcal{V}^{-1}$, where $\mathcal{V}$ is the overall volume of $\tilde{X}_4$,
\be
\tilde{J} = \phi^i \omega_i + L^{\alpha} \omega_{\alpha}.
\label{kahler}
\ee 
$\omega_{\alpha}$ are two-forms other than $\omega_i$.

Since the chiral multiplets carrying weight ${\bf w}$ can be interpreted in terms of M2-branes wrapping an effective curve $\Sigma_{{\bf w}}$, it follows that the Dynkin label is the coupling between the M2-branes and the $U(1)$ gauge field, i.e. the Dynkin labels of the chiral multiplets can be geometrically characterized by 
\be
q_i = \int_{\Sigma_{{\bf w}}} \omega_i = \Sigma_{{\bf w}}\cdot D_i \,.
\label{dynkin}
\ee
Note that the geometric Dynkin labels are the negative of the gauge theoretic Dynkin labels. One may understand this by considering the intersection between curves corresponding to the simple roots and the (the dual) divisors $D_i$, which intersect in minus the Cartan matrix. 
Hence, in order to make contact with the gauge theories, we consider the negative of the geometric Dynkin labels. 

By combining the two equations \eqref{kahler} and \eqref{dynkin}, the negative of the relative K\"ahler cone exactly determines the classical phase structures of the classical Coulomb branch in the three-dimensional $\cN=2$ gauge theories. Hence,  the relative Mori cone contains the information about which weight satisfies $\phi \cdot {\bf w} > 0$ or $\phi \cdot {\bf w} > 0$. If a curve $\Sigma_{{\bf w}}$ is inside the relative Mori cone, then, the corresponding weight should satisfy $\phi \cdot {\bf w} < 0$. 
The same structure can be seen in the correspondence between the Coulomb branch of five-dimensional supersymmetric gauge theories and the negative of the relative K\"ahler cone of resolved Calabi-Yau threefolds \cite{Intriligator:1997pq},
where novel physical phenomena are found at the boundary of the (extended) K\"ahler cone \cite{Witten:1996qb, Morrison:1996xf}.

The relative Mori cone in resolved Calabi-Yau fourfolds can be determined from the structure of the fiber in higher codimension. Recall that in codimension 1 the resolution gives rise to so-called Cartan divisors, which are obtained by fibering the resolution $\mathbb{P}^1$s, labeled by roots, over $S$. 
As noted in \cite{MS} one way to study the structure at the codimension 2 loci is to follow the Cartan divisors to the codimension 2 loci. Some of the Cartan divisors, when restricted to the matter loci will become reducible and {correspond to the matter surfaces, which are $\mathbb{P}^1$ fibrations over the matter curves inside $S$}.

From these we can now extract the relative Mori cone. Note that this contains holomorphic curves that shrink in the singular limit, so that the curves $\Sigma_{{\bf w}}$ appearing in the resolution along the codimension 2 loci should be contained in the relative Mori cone. Furthermore, any non-negative linear combination of the $\Sigma_{{\bf w}}$ will also be contained in the relative Mori cone. 
%%???The linear combination of the $\Sigma_{{\bf w}}$'s is labeled by the same linear combination of the weights ${\bf w}$'s corresponding to the $\Sigma_{{\bf w}}$'s. 
Finally, along codimension 3 loci, curves $\Sigma_{{\bf w}}$ can intersect, and thereby the phase structure of different representations mix. For instance the codimension 3 locus corresponding to a coupling ${\bf 10} \times {\bf 10}\times {\bf 5}$ will mix the phases of the ${\bf 10}$ and ${\bf 5}$ matter fields. This is analogous to the compatibility condition that we discussed in the gauge theoretic phases. 
Combining the codimension 2 and 3 information, we can then recover the 
corresponding phase of the three-dimensional $\cN=2$ gauge theories.

%%%%%%%%%%%%%%%%%%%%%%%%%%%%%%%%%%%%
%%%%%%%%%%%%%%%%%%%%%%%%%%%%%%%%%%%%
\section{Geometric Phases from Toric Resolutions}
\label{sec:toric}

There are two types of methods -- toric and algebraic -- that we will use to resolve {the singular} Calabi-Yau fourfold. 
First consider toric resolutions of elliptically fibered Calabi-Yau fourfolds $X_4$ with base $B_3$.  
Let $z$ be a homogeneous coordinate %on the base,
whose vanishing defines a component $S$ of the discriminant of the elliptic fibration, with an $A_4$ singularity. 
Such an elliptically fibered Calabi-Yau fourfold can be globally written in Tate form \cite{Bershadsky:1996nh, Katz:2011qp}
\be
\label{Tate}
P_T:
\qquad  y^2 +b_1 w x y+b_3z^2  w^3y = x^3+b_2 z w^2x^2+b_4 z^3 w^4x+b_6 z^5w^6\,.
\ee
which is a hypersurface in the auxiliary five-fold which is a $\mathbb{P}^2_{1,2,3}$ bundle over $B_3$.
%Here, $w, x, y$ are sections of $\mathcal{O}(\sigma)$, $\mathcal{O}(2\sigma+ 2c_1)$ and  $\mathcal{O}(3\sigma+ 3c_1)$, where $\sigma$ is the degree one section of the $W\mathbb{P}^2_{1,2,3}$ fiber, and  %For most purpose we will set $w=1$. 
The classes of the sections appearing in the Tate form are 
\be
[b_n] =(nc_1- i_nS) \,,\qquad (i_n)_n = (0,1,2,3,5) \,,
\ee
where $c_1$ is the pullback of the first Chern class of the base $B_3$.
There are two loci of codimension 2 enhancement of the symmetry
\be\label{Codim2Loci}
\ba
D_5:\qquad &b_1=z= 0\cr
A_5:\qquad & P\equiv b_2b^2_3 + b_1 (b_1b_6 - b_3b_4) =z=0 \,, 
\ea
\ee
where the ${\bf 10}$ and $\bar{{\bf 5}}$ matter is localized.

%%%%%%%%%%

\subsection{{Toric Resolutions}}

The toric resolution of this geometry was obtained in \cite{Candelas:1996su, Bershadsky:1996nh, Candelas:1997eh} and can be summarized by
\be
(x, y, z) \rightarrow (xe_1e_4e_2^2e_3^2, \, ye_1e_4^2e_2^2e_3^3,\,  e_0 e_1e_2e_3e_4), \label{resolution}
\ee
where $\{ e_i = 0 \}, i=1,\cdots, 4$ stands for the corresponding blow up divisors. The divisor $\{e_0 = 0 \}$ will correspond to the extended Dynkin node of the extended Dynkin diagram of $A_4$. The proper transformation of the resolved Tate form is 
\be
\ba
\tilde{P}_T:\qquad  &y^2e_3e_4 + b_1xyw + b_3 yw^3e_0^2e_1e_4 \cr
&\qquad = x^3e_1e_2^2e_3 + b_2 x^2w^2e_0e_1e_2 + b_4 xw^4e_0^3e_1^2e_2e_4 +b_6 w^6e_0^5e_1^3e_2e_4^2  \,.
\label{rTate}
\ea\ee
There are in fact various routes to arrive at \eqref{resolution} and each inequivalent route corresponds to a different resolution. In appendix \ref{app:ToricAlg} the algebraic resolutions that give rise to these are constructed.

In terms of toric data the resolution was obtained in \cite{Candelas:1996su, Candelas:1997eh}. The homogeneous coordinates $x, y, w, e_0, e_1, e_2, e_3, e_4$ are specified by the points in a six-dimensional lattice
\begin{eqnarray}
\begin{array}{c|l}
\hbox{Section} & \hbox{Toric Vector}\\
\hline
x & (-1,  0,   \vec{0})\\
%\hline
y & (0, -1, \vec{0}) \\
%\hline
w & (2, 3,  \vec{0}) \\
%\hline
e_0 & (2, 3, \vec{v}) \\
%\hline
e_1 & (1, 2, \vec{v}) \\
%\hline
e_2 & (0, 1, \vec{v}) \\
%\hline
e_3 & (0, 0, \vec{v}) \\
%\hline
e_4 & (1, 1, \vec{v}) \\
%\hline 
\end{array}
\label{toric}
\end{eqnarray} 
where $\vec{0}$ denotes the three-dimensional zero vector and the $\vec{v}$ is a three-dimensional vector in the lattice. The resolved Calabi-Yau fourfold $\tilde{X}_4$ is a hypersurface in the toric ambient space \eqref{toric}\footnote{Certainly, we have to specify more vertices to fully identify the whole toric ambient space. However, the toric resolution of the $A_4$ singularity can be determined by considering the triangulation of a fan specified by the points written in \eqref{toric}}. Each of the new sections $e_i=0$ corresponds to a divisor in the resolved Tate form, which we will denote by $D_{-\alpha_i}$, the Cartan divisors.

Let us see each resolution phase in more detail. The resolution structure was essentially studied in \cite{Krause:2011xj} for the  model with additional $U(1)$ and in \cite{Grimm:2011fx} for examples with and without a $U(1)$. In order to find the resolution structure, it is enough to look at the Stanley-Reisner ideal, which is characterized by a set of coordinates which do not vanish simultaneously. The Stanley-Reisner ideal is different for each triangulation. The common part of the Stanley-Reisner ideal of \eqref{toric} is 
\be
\left\{ xyw, ye_1, ye_2, we_3, xe_4, e_0e_2, xe_0e_3, xe_1e_3, ye_0e_3, we_1e_4, we_2e_4, xye_0 \right\}.\label{SR1}
\ee
There are further elements in the Stanley-Reisner ideal, which {depend} on the triangulations:
\bea
\left\{
\begin{array}{c}
ye_0\\
w e_4
\end{array}
\right\} \times \left\{
\begin{array}{c}
xe_0, xe_1\\
xe_0, we_2\\
we_1, we_2
\end{array}
\right\} \times \left\{
\begin{array}{c}
e_0e_3, e_1e_3\\
e_0e_3, e_2e_4\\
e_1e_4, e_2e_4
\end{array}
\right\}.
\label{SR2}
\eea
Hence, we have eighteen triangulations in total. 
However, eq.~\eqref{SR1} and \eqref{SR2} are the Stanley-Reisner ideal from the triangulation of the toric {\it ambient} space. The number of triangulations of the hypersurface is in general less than the number of the triangulations of the toric ambient space. In fact, each element in the first two columns of \eqref{SR2} does not vanish inside the Calabi-Yau fourfold hypersurface. For example, $y=0, e_0=0$ is not compatible with \eqref{rTate} due to the common Stanley-Reisner ideal \eqref{SR1}. Hence, the phases of the resolved Calabi-Yau fourfolds are characterized by 
\be\label{tSR}
\ba
\hbox{Toric Resolution I}: \qquad &e_0e_3,\,  e_1e_3\cr 
\hbox{Toric Resolution II}:\qquad & e_0e_3, e_2e_4\cr
\hbox{Toric Resolution III}: \qquad &e_1e_4, e_2e_4 \,.
\ea
\ee
Namely, we have three resolved phases. 
With the defining equation \eqref{rTate} and the Stanley-Reisner ideal {generated by} \eqref{SR1} and \eqref{tSR}, one can determine the Dynkin diagram and the weights corresponding to their nodes at all the singularity loci for all three phases. In particular, we will now consider the splitting of the Cartan divisors, defined by $e_i=0$ along the codimension 2 loci, and thereby we can determine which weights of $SU(5)$ correspond to effective curves inside the relative Mori cone.

Before going to the results, let us point out that there are two choices for the order of the Cartan divisors:
\bea
\hbox{Cartans 1}:\  &
D_{-\alpha_1}: \  e_1=0 \,,\quad 
D_{-\alpha_2}: \ e_2=0 \,,\quad 
D_{-\alpha_3}: \  e_3=0 \,,\quad 
D_{-\alpha_4}: \  e_4=0  \,, \label{Cartan1} \\
\hbox{Cartans 2}:\  &
D_{-\alpha_1}: \  e_4=0 \,,\quad 
D_{-\alpha_2}: \  e_3=0 \,,\quad 
D_{-\alpha_3}: \ e_2=0 \,,\quad 
D_{-\alpha_4}: \ e_1=0 \,.  \label{Cartan2} 
\eea
The two choices are related by the $\Z_2$ automorphism of the affine $A_4$ Dynkin diagram, i.e. the outer automorphism of $A_4$. The resolution of the $A_4$ singularity in codimension 1 cannot distinguish the two choices since the geometry only knows the extended $A_4$ Dynkin diagram, however it makes a difference in codimension 2. We denote the two choices by choice 1 for \eqref{Cartan1} and choice 2 for \eqref{Cartan2}.

%%%%%%%%%%%%%%%%%%%%

\begin{figure}[tb]
\begin{center}
\includegraphics[width=80mm]{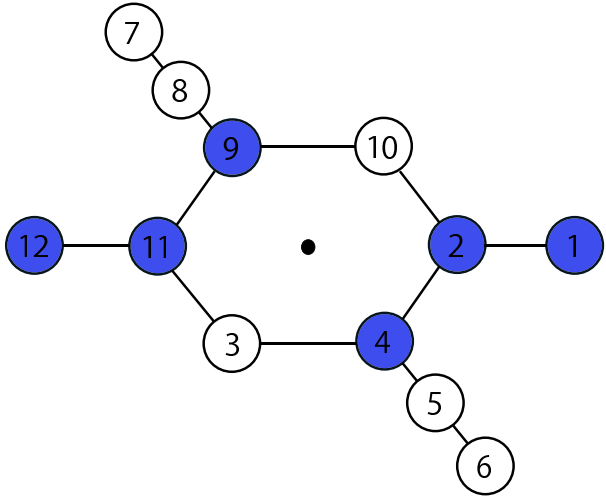}
\end{center}
\caption{Phase diagram, with blue nodes representing the phases that have a realization in terms of toric resolutions of the singularity. }
\label{fig:ToricPhases}
\end{figure}

%%%%%%%%%%%%%%%%%%%%

\subsection{{Phases from Toric Resolutions}}

Let us move on to the results of the resolution structure from the toric blow ups. We label the three resolutions by Toric Resolutions I, II and III  corresponding to the Stanley-Reisner ideal as detailed in (\ref{tSR}), and in addition there is a choice of Cartan divisors, as in (\ref{Cartan1}, \ref{Cartan2}). We will sometimes refer to one choice in terms of I$.1$, I$.2$ etc. 
We only write down explicitly the weights for the choice 1 of the Cartans. The weights for the choice 2 follow by the $\Z_2$ transformation of the weights in choice 1. 

The generic fiber in codimension 1 is characterized by the vanishing of $e_i$ inside the resolved Tate form 
\be
D_{-\alpha_i}:\qquad \tilde{P}_T = e_i =0 \,.
\ee
and are labeled by the simple roots of $A_4$. These are the so-called Cartan divisors. 
We now consider how these split along the codimension 2 loci (\ref{Codim2Loci}), as in \cite{MS}. 

First consider {\it Toric Resolution I} with choice 1 of the Cartans.  Along the {\bf 10} matter locus the only Cartan divisors that become reducible are 
\be
b_1=0:\qquad 
\ba
D_{-\alpha_2} &\quad \rightarrow\quad  (0,1,-1,1,-1) + (0, 0, -1, 0, 1)\cr
D_{-\alpha_4} &\quad \rightarrow \quad (1,-2,1, 0, 0) + (0,1,-1,1,-1) + (0,1,0,0,-1) \,.
\label{I-10}
\ea
\ee
The weights appearing on the RHS are computed from the intersections of the irreducible curve components of $D_{-
\alpha_2}.(b_1=0)$ with all the Cartan divisors, including the one for the extended node $D_{-\alpha_0}$, which is the first entry.
Along the ${\bf\bar{5}}$ matter locus the only reducible Cartan divisor is
\be
P=0:\qquad D_{-\alpha_3} \quad \rightarrow \quad (0,0,1,-1,0) + (0,0,0,-1,1)\,.
\label{I-5}
\ee
From these splittings, we now can determine the generators of the relative Mori cone: since all the curves\footnote{Although $D_{-\alpha_i}.(b_1=0)$ or $D_{-\alpha_i}.(P=0)$ are surfaces in the resolved Calabi-Yau fourfold $\tilde{X}_4$, one can generically make them a curve by intersecting with a divisor which is a pull-back of a divisor in $S$ transversally intersecting with $b_1=0$ or $P=0$ at a point in $S$.} appearing in \eqref{I-10} and \eqref{I-5} and also the irreducible curves corresponding to the negative of the simple roots are inside the blow up divisors, they vanish in the singular limit. Also, they are holomorphic curves since they can be explicitly expressed by  holomorphic defining equations. Therefore, all the elements of \eqref{I-10} and \eqref{I-5}, and also the irreducible curves corresponding to the negative of the simple roots that remain irreducible are inside the relative Mori cone. A choice of four weights or roots out of these which generate  the relative Mori cone in this phase are
\be
\hbox{Toric Resolution I.1} \qquad (2,-1,0,0),\;\; (-1,1,-1,1),\;\;(0,-1,1,0),\;\;(0,1,0,-1).
\ee
This is exactly the same set appearing in Table \ref{tb:generators} as the phase 9. Hence, the Toric Resolution I.1 corresponds to the phase 9 on the gauge theory side. Similarly, making the choice 2 for the Cartans, the Toric Resolution I.2 corresponds to the phase 4.
%The weights appearing in the Dynkin diagram for the phase I of the choice 1 is 
%\bea
%\begin{array}{|c|c|}
%\hline
%(i) & (ii)\\
%\hline
%(1,0,0,1) & (1,0,0,1)\\
%\hline
%(-2,1,0,0) & (-2,1,0,0)\\
%(1,-1,1,-1) & (1,-1,1,-1)\\
%(0,1,-1,0) & (0,1,-1,0) \times 2\\
%(0,-1,0,1) & (0,-1,0,1)\\
%(0,0,-1,1) & (1,0,0,-1)\\
%\hline
%\end{array}
%\label{phaseI}
%\eea 
%The first row represents the nodes which do not vanish in the singular limit. The expression $(0,1,-1,0) \times 2$ means that there are two nodes which correspond to the weight $(0,1,-1,0)$. The nodes of (i) in \eqref{phaseI} forms the Dynkin diagram of $E_6$ and the nodes of (ii) in \eqref{phaseI} forms the extended Dynkin diagram of $D_6$ as first noted in \cite{Esole:2011sm}. 

One can perform the same analysis for the Toric Resolution II.1. The Cartan divisors that become reducible along the ${{\bf 10}}$ matter locus $b_1= 0$ are 
\be\label{II-10}
b_1=0:\qquad \ba
D_{-\alpha_1} & \quad \rightarrow\quad (0,-1,1,-1,1) + (1,-1,0,1,-1)\cr
D_{-\alpha_3} & \quad \rightarrow\quad (0,-1,1,-1,1) + (0,1,0,-1,0)\cr
D_{-\alpha_4} & \quad \rightarrow\quad (1,-1,0,1,-1) + (0,1,0,0,-1) \,,
\ea
\ee
and along the ${\bar{\bf{5}}}$  matter locus $P=0$
\be\label{II-5}
P=0: \qquad D_{-\alpha_3} \quad \rightarrow\quad  (0,0,1,-1,0) + (0,0,0.-1,1)\,.
\ee
Again, the generators of the relative Mori cone follow from these decompositions as 
\be
\hbox{Toric Resolution II.1}:\qquad (-1,2,-1,0),\;\; (1,-1,1,-1),\;\; (1,0,-1,1),\;\;(-1,0,0,1) \,.
\ee
They are the generators of the cone for the phase 11 in the Table \ref{tb:generators}. Accordingly, the  Toric Resolution II.2
corresponds to the phase 2.
%The weighs appearing in the Dynkin diagram of the phase II of the choice 1 is 
%\bea
%\begin{array}{|c|c|}
%\hline
%(i) & (ii)\\
%\hline
%(1,0,0,1) & (1,0,0,1)\\
%\hline
%(1,-2,1, 0) & (1,-2,1, 0)\\
%(-1,1,-1,1) & (-1,1,-1,1)\\
%(-1,0,1,-1) & (-1,0,1,-1) \\
%(1,0,0,-1) & (1,0,0,-1)\\
%(0,0,-1,1) & (0,-1,1,0) \times 2 \\
%\hline
%\end{array}
%\label{phaseII}
%\eea 
%In this case, the Dynkin diagram coming from the resolution of the codimension-three singularity of type (i) forms the so-called $T^{-}_{3,3,3}$ diagram.  

Finally, let us consider the Toric Resolution III.1, where along the ${\bf 10}$ matter locus the splitting of the Cartans is
\be\label{III-10}
b_1=0:\qquad 
\ba
D_{-\alpha_0} & \quad \rightarrow\quad (-1,1,0,-1,1) + (-1,0,0,1,0)\cr 
D_{-\alpha_3} & \quad \rightarrow\quad (-1,1,0,-1,1) + (1,-2,1,0,0) + (0,1,0,-1,0)\,, 
\ea
\ee
and the others are irreducible. %Finally, The weighs appearing in the Dynkin diagram of the phase III for the choice 3 is 
%\bea
%\begin{array}{|c|c|}
%\hline
%(i) & (ii)\\
%\hline
%(0,0,1,0) & (0,0,1,0)\\
%\hline
%(-2,1,0, 0) & (-2,1,0, 0)\\
%(1,-2,1,0) & (1,-2,1,0)\\
%(0,0,1,-2) & (0,0,1,-2) \\
%(1,0,-1,1) & (1,0,-1,1)\\
%\hline
%\end{array}
%\label{phaseIII}
%\eea 
%The shape of the Dynkin diagrams is essentially the same as the ones of the phase I. 
Here, there is a clear difference between the case III and the cases I, II. Namely, in the Toric Resolution III.1, Cartan divisor corresponding to the extended node of the extended $A_4$ Dynkin diagram  splits  along a higher codimension locus. 

Let us see this phenomenon explicitly by focusing on the ${\bf 10}$ matter curve $b_1 = 0$. The Cartan divisor $D_{-
\alpha_0}$ restricts to 
\be
\{ e_0 = 0 \} \cap \{ e_3(y^2e_4 - x^3e_1e_2^2) = 0\} \cap \{ b_1 = 0 \}.
\ee
Hence, if $e_0$ and $e_3$ can simultaneously vanish the extended node can split into two components. In the phase I and phase II, $e_0e_3$ is inside the Stanley-Reisner ideal \eqref{tSR} but it is not inside the Stanley-Reisner ideal for the phase III. Hence, the extended node of the extended $A_4$ Dynkin digram does split in the phase III. $\{e_0 = 0\} \cap \{e_3 = 0\} \cap \{b_1=0 \}$ corresponds to the weight $(-1,1,0,-1,1)$ and vanishes in the singular limit. Hence, one has to take into account that the weight $(-1,0,0,1,0)$ is not inside the relative Mori cone. %This characteristic becomes important when one compares the toric resolution with the small resolution in the later section. 

To complete the analysis of this case, note that the decomposition along the ${\bar{{\bf 5}}}$ matter for the case III.1 is
\be
P=0:\qquad D_{-\alpha_3} \quad \rightarrow\quad (0,0,1,-1,0) + (0,0,0.-1,1)\,,\label{III-5}
\ee
and the others are irreducible. Therefore, the generators of the relative Mori cone for the Toric Resolution III.1 are 
\be
\hbox{Toric Resolution III.1}:\qquad (-2,1,0,0),\;\;(-1,2,-1,0),\;\;(0,0,-1,2),\;\;(-1,0,1,-1) \,.
\ee
In this case, those weights appear as the generators of the cone for the phase 12 in table \ref{tb:generators}. Likewise, {it follows that,} after applying the $\Z_2$ automorphism, the resolution III.2 corresponds to the gauge theory phase 1.

To summarize, the relation between the geometric resolutions and the gauge theory phases in the case of the toric blow ups is 
\bea
\begin{array}{c|c|c}
\text{Toric Resolution} & \text{Cartan choice}  & \text{Gauge Theory Phase}\\
\hline
I & 1&  9\\
I & 2&  4\\
 \hline
II & 1&  11 \\
II & 2&  2\\
 \hline
III & 1&  12\\
III  & 2& 1\\
\end{array}
\eea
The toric resolutions do not reproduce all the phases of the gauge theory. It is therefore key to consider also algebraic resolutions, which will lead to the completion of the picture in figure \ref{fig:flop}. This will be discussed in the next two sections. 

%%%%%%%%%%%%%%%%%%%%%%%%%%%%%%%%%%%%
%%%%%%%%%%%%%%%%%%%%%%%%%%%%%%%%%%%%

\section{Geometric Phases from Algebraic Resolutions}
\label{sec:AlgRes}

From the toric point of view we did not realize all phases that are seen in the gauge theory. Alternatively we can consider algebraic resolutions of the singularity. We will show that these generate additional phases, and that we can use them as  a starting point to apply flops to generate {the complete phase diagram.}

%%%%%%%%%%%%%%%%%%%%%%%%%%%%%%%%%%%%
\subsection{Resolution in Codimension 1}
\label{subsec:Codim1}

The starting point for the algebraic resolution is the Tate form for $SU(5)$ \cite{Bershadsky:1996nh, Katz:2011qp}
\be
\label{TateAlg}
 w y^2 +b_1 w x y+b_3z^2  wy = x^3+b_2 z wx^2+b_4 z^3 w^2x+b_6 z^5w^3\,,
\ee
which, for the purpose of the algebraic resolutions, we construct as a hypersurface in the auxiliary five-fold which is a $\mathbb{P}^2$ bundle over $B_3$
\be
X_5= \mathbb{P} (\mathcal{O}\oplus K^{-2}_{B_3} \oplus K^{-3}_{B_3} )\,.
\ee
Here, $w, x, y$ are sections of $\mathcal{O}(\sigma)$, $\mathcal{O}(\sigma+ 2c_1)$ and  $\mathcal{O}(\sigma+ 3c_1)$, where $\sigma$ is the hyperplane section of the $\mathbb{P}^2$ fiber, and $c_1$ is the pullback of the first Chern class of the base $B_3$. For most purposes we will set $w=1$.

The Tate form of $SU(5)$ is 
\begin{equation}
y^2 +b_1 x y+b_3 \zeta _0^2 y = x^3+b_2 \zeta _0 x^2+b_4 \zeta _0^3 x+b_6 \zeta _0^5 \,,
\end{equation}
where the $SU(5)$ singular fiber is located along $\zeta_0=0$. 
The singularity in codimension 1 can be resolved by two blowups
\begin{equation}
\ba
&(x, y, \zeta_0 ; \zeta_1) \cr
& (x, y, \zeta_{1};\zeta_2) \,,
\ea
\end{equation}
where the notation, as in \cite{Lawrie:2012gg}, indicates 
\begin{equation}
(x, y, \zeta_0 ; \zeta_1):\qquad x\rightarrow x \zeta_1 \,,\quad y\rightarrow y \zeta_1 \,,\quad \zeta_0 \rightarrow \zeta_0 \zeta_1 \,,
\end{equation}
where the new sections satisfy projectivity $[x, y, \zeta_0]$. 
The proper transform of the resulting codimension 1 resolved space is 
\begin{equation}\label{Res1}
y \left(y+ b_1 x+b_3 \zeta _1 \zeta _0^2\right) =
\zeta _1 \zeta _2 \left(b_2 x^2 \zeta _0+\zeta _1 \zeta _0^3 \left(b_6 \zeta _1 \zeta _0^2+b_4 x\right)+\zeta _2 x^3\right) \,.
\end{equation}
We will abbreviate this often by
\begin{equation}\label{Res1Short}
y Y = \zeta _1 \zeta _2 C \,.
\end{equation}
The sections have to satisfy the following projectivity relations
\begin{equation}
	\ba
		&{[x\zeta_2, y\zeta_2, \zeta_0] }\cr
		&[x, y, \zeta_1] \,.
	\ea
\end{equation}
The sections after the two blowups have the following classes
\begin{equation}
\begin{array}{l|l}
\hbox{Section} & \hbox{Class} \cr\hline
x & \sigma+2c_1 - E_1 - E_2 \cr
y& \sigma+ 3 c_1 - E_1 - E_2 \cr
\zeta_0 & S-E_1\cr
\zeta_1 & E_1- E_2 \cr
\zeta_2 & E_2 \cr
\end{array}
\end{equation}

The space (\ref{Res1}) is resolved in codimension 1, as can be readily checked. 
The exceptional sections $\zeta_1$ and $\zeta_2$ are reducible and give rise to four irreducible exceptional divisors. However, the space is still singular in higher codimension. There are various ways to resolve this space, which we will now consider. 
Note first that (\ref{Res1Short}) has the general structure of a binomial geometry 
\begin{equation}\label{BinGeo}
  v_1 v_2 = u_1 u_2 u_3 \,.
\end{equation}
Here
\begin{equation}
\ba
v_1&= y \cr
v_2 & =y+ b_1 x+b_3 \zeta _1 \zeta _0^2\cr
u_1 &=  \zeta_1 \cr
u_2 &= \zeta_2 \cr
u_3&=b_2 x^2 \zeta _0+\zeta _1 \zeta _0^3 \left(b_6 \zeta _1 \zeta _0^2+b_4 x\right)+\zeta _2 x^3 \,.
\ea
\end{equation}
There are several choices of small resolutions that will resolve the space fully. 
Denote the small resolutions by 
\begin{equation}\label{SMijkl}
((i,j), (k,l)):\qquad (v_i, u_j; \delta_1) \quad \hbox{and } \quad (v_k, u_l; \delta_2) \,.
\end{equation}
This corresponds to the small resolutions where $[v_i, u_j]$ and $[v_k, u_l]$ form each a new $\mathbb{P}^1$ with exceptional sections $\delta_1$ and $\delta_2$ respectively.

Note, that after the small resolutions, we can read off, as in \cite{MS}, the sections for the exceptional divisors from the transformation of $\zeta_0$. For instance if the small resolution only involves $ijkl\in\{1,2\}$ we always have 
\be\label{zeta0trans}
\zeta_0 \quad \longrightarrow\quad \zeta_0 \zeta_1 \zeta_2 \delta_1 \delta_2 \,,
\ee
which confirms that $\zeta_i$ and $\delta_i$ are the exceptional sections.

%%%%%%%%%%%%%%%%%%%%

\begin{figure}[tb]
\begin{center}
\includegraphics[width=80mm]{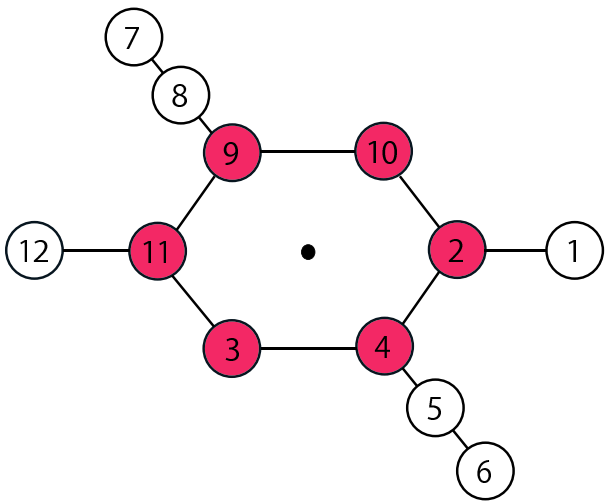}
\end{center}
\caption{Phase diagram, where the red dots label the phases that have a realization in terms of small resolutions using direct algebraic resolution of the singularity defined in (\ref{SMijkl}). }
\label{fig:AlgPhases}
\end{figure}

%%%%%%%%%%%%%%%%%%%%

%%%%%%%%%%%%%%%%%%%%%%%%%%%%%%%%%%%%

\subsection{Network of Small Resolutions}

Each of the small resolutions  (\ref{SMijkl}) yields a specific splitting along the {\bf 10} and ${\bf \bar{5}}$ curves, which together with the codimension 3 Yukawa couplings, can then be identified with a gauge theory phase by reading off the generators of the relative Mori cone. Before getting to the details, we summarize the small resolutions and the corresponding phases:
\begin{equation}\label{AlgResSum}
\begin{array}{c|c}
\hbox{ Algebraic Resolution } & \hbox{ Phases }  \cr
\hline
((1,1),(1,2)) & 4 \hbox{ and }9\cr
%((1,2),(1,3)) & 4 \hbox{ and }9\cr
%((1,1),(2,3)) & 4 \hbox{ and }9\cr
((2,1),(2,2)) & 4 \hbox{ and }9\cr\hline
((2,2),(1,1)) & 3 \hbox{ and }10\cr
((1,1),(2,2)) & 3 \hbox{ and }10\cr
((2,1),(1,2)) & 3 \hbox{ and }10\cr
((1,2),(2,1)) & 3 \hbox{ and }10\cr\hline
%((1,3),(1,1)) & 2 \hbox{ and }11\cr
((1,2),(1,1)) & 2 \hbox{ and }11\cr
%((1,2),(2,3)) & 2 \hbox{ and }11\cr
((2,2),(2,1)) & 2 \hbox{ and }11
\end{array}
\end{equation}

Note that each entry gives two phases, which are related by a simple reordering of the Cartan divisors, explicitly, there is a $\mathbb{Z}_2$ choice 
\be \label{Z2Choice}
\mathbb{Z}_2:\qquad \alpha_1 \leftrightarrow \alpha_4\,,\qquad \alpha_2 \leftrightarrow \alpha_3\,.
\ee 
Some of  the small resolutions  in table (\ref{SMijkl}) have appeared before in the literature. The case $((1,1),(1,2))$ was discussed in \cite{Lawrie:2012gg}, and corresponds to phase 9. As this will be used later on for one of the flops we summarized it in appendix \ref{app:1112}. The small resolution $((1,1), (2,2))$ was done in \cite{MS} and corresponds to phase 10. 
The algebraic resolutions (\ref{AlgResSum}), including the identification of the corresponding phases, are discussed in detail in appendix \ref{app:AlgRes}.

Let us add a remark concerning the relation to \cite{Esole:2011sm}. In that paper the binomial geometry (\ref{BinGeo}) was resolved in higher codimension by toric methods. It is clear, and already noticed in \cite{MS}, that these correspond to the small resolutions of the type
\be\label{EYOnes}
((i,k), (j,l)) \,,\qquad i \not= j \,.
\ee
There are 6 such small resolutions and they agree with the ones in \cite{Esole:2011sm}. The remaining algebraic ones, i.e. $i=j$, are equivalent to these 6 resolutions, however, are generically easier to compute with. In appendix \ref{subsec:Consistency} we give a detailed account why the small resolutions of this type with $i=j$ are indeed consistent resolutions, in particular, why they are isomorphisms away from the singular loci. 

Also, as we will show in the next section, the latter are the starting points for the flops along matter curves\footnote{Of course in principle the flops can be performed from the resolutions in (\ref{EYOnes}), however, as is clear from the resolution in \cite{MS}, the ones with $i\not=j$ generically have additional relations between the coordinates, which makes the computations unnecessarily involved.}

We did not include any small resolutions along the $u_3=0$ component in table (\ref{AlgResSum}), since
all {24 small resolutions of the type $((i, j), (k, l))$ where $j \neq l$}\footnote{{The case where $j = l$ is uninteresting as the second small resolution does not resolve any higher codimension singularities. In the case where $k = l = 3$ one needs to do more resolutions to resolve the space, and these additional resolutions govern the phase realized.}} defined in (\ref{SMijkl}) realize one of the 6 phases that appear already in table \ref{AlgResSum}. 
We have considered the resolutions involving the additional section $u_3$ (or $C$, as it is denoted in (\ref{Res1Short})), which result in the phases 9, 10, or 11, and thus do not add any new phases. However, there is an interesting point here: 
consider for instance $((1, 1), (1, 3))$. In this case, the exceptional divisor $\zeta_2=0$ is in fact reducible, given by $v_1 v_2=0$.  This point is also clear when considering the transformation of $\zeta_0$ under the resolution. Unlike (\ref{zeta0trans}), which is for the cases that do not involve $u_3$, in the case $((1, 1), (1, 3))$ we would have
$\zeta_0  \rightarrow \zeta_0 \zeta_1\zeta_2 \delta_1$, 
but no $\delta_2$. Nevertheless the resolution reproduces a $A_4$ fiber, because $\zeta_2$ is not irreducible, and each irreducible components corresponds to a Cartan divisor. 
In fact, in general, performing two small resolutions, where one is 
along the $u_3$ component, one of the exceptional sections is always left reducible, i.e. does not correspond to 
{a Cartier divisor}. This can be remedied by an additional small resolution, in this case along $(1,2)$, with a new exceptional section, $\hat\delta$ and thereby obtaining 
\be\label{zeta0Extra}
\zeta_0 \quad \longrightarrow\quad \zeta_0 \zeta_1\zeta_2 \delta_1\hat\delta\,.
\ee
The additional small resolution does not change the phase, but makes the structure of exceptional divisors more transparent. 
 Similar situations arise in the case of resolutions of general ADE singularities in higher codimension as discussed in \cite{Lawrie:2012gg}. 

In the following we will always consider the case where the Cartan        
divisors are Cartier, in particular the zero locus of the exceptional  
sections is irreducible. To achieve this, in the case when $u_3$ is involved, one can do an additional
small resolution. An example of this kind is given in appendix \ref{app:smallres3} and in
 section \ref{subsec:Phase11}.

%%%%%%%%%%%%%%%%

\subsection{Small Resolution ((1,3),(1,1),(1,2)), Phase 2 and 11}
\label{subsec:Phase11}

To illustrate the point that we made in the last section about irreducibility of the exceptional divisors in the case when $u_3$ is used in the small resolutions, consider 
the case $((1,3), (1,1))$, which we will show corresponds to phases 2 and 11. 
The required small resolutions are explicitly
\begin{equation}
\ba
& (y, C; \delta_1) \cr
&(y, \zeta_1; \delta_2) \,.
\ea\end{equation}
Recall the notation (\ref{Res1Short}) for $C$, 
which is not irreducible, so that in fact we obtain an additional relation
\begin{equation}\label{Phase11Extra}
C \delta_1
 = b_2 x^2 \zeta _0+\delta_2 \zeta _1 \zeta _0^3 \left(b_6 \delta_2 \zeta _1 \zeta _0^2+b_4 x\right)+\zeta _2 x^3 \,.
\end{equation}
The fully resolved geometry takes the simple form
\begin{equation}
y \left(b_3\delta_2 \zeta _1 \zeta _0^2+b_1 x+\delta _1 \delta _2 y\right)=  \zeta _1 \zeta _2 C \,.
\end{equation}

It is clear from this equation that $\zeta_2=0$ is however not irreducible, as there is no projective relation that prevents $\zeta_2=0$ and $y=0$. This is precisely the situation alluded to in the last section. To make all the exceptional divisors irreducible, consider an 
extra small resolution along $\zeta_2=y=0$, i.e. $(1,2)$
 \begin{equation}
(y, \zeta_2; \delta_5) \,.
\end{equation}
In particular, the exceptional divisors are now all Cartier, with the exceptional sections given by $\zeta_0$, $\delta_2$, $\zeta_1$, $\zeta_2$ and $\delta_5$, as in (\ref{zeta0Extra}).

To determine the gauge theory phase corresponding to this small resolution, we need to analyze the splitting of the $A_4$ fiber along the codimension 2 (and 3) loci. Consider first the {\bf 10} matter locus, i.e. $b_1=0$, along which we find 
\begin{equation}
\ba
\delta_2=0:\qquad -\alpha_1=({1, -2, 1, 0, 0})& \quad \longrightarrow \quad ( {1, -1, 0, 1, -1})+( {0, -1, 1, -1, 1})\cr
\zeta_2=0:\qquad -\alpha_3=({0, 0, 1, -2, 1})  & \quad \longrightarrow \quad ({0, -1, 1, -1, 1} )+( {0, 1, 0, -1, 0}) \cr
\zeta_1=0:\qquad -\alpha_4=  ({1, 0, 0, 1, -2})& \quad \longrightarrow \quad ( {1, -1, 0, 1, -1})+( {0, 1, 0, 0, -1})\,,
\ea\end{equation}
whereas $D_{-\alpha_0}$ and $D_{-\alpha_2}$ corresponding to $\zeta_0=0$ and $\delta_5=0$ respectively, stay irreducible. 
Along the ${\bf \bar{5}}$ matter locus $P=0$ the only Cartan divisor that splits is
\begin{equation}
\ba
%-\alpha_2= ( {0, 1, -2, 1, 0}) &\quad\rightarrow \quad( {0, 0, -1, 1, 0})+( {0, 1, -1, 0, 0})\cr
-\alpha_3=
( {0, 0, 1, -2, 1}) &\quad\longrightarrow \quad ( {0, 0, 1, -1, 0})+( {0, 0, 0, -1, 1}) \,.
\ea
\end{equation}

Combining the information from these splittings into a basis for the Mori cone in this small resolution, we can identify it with phase 11. The detailed splittings of matter in the remaining algebraic resolutions are summarized in appendix \ref{app:AlgRes}. By reversal of assignment of the roots to the divisors, i.e. under the $\mathbb{Z}_2$ automorphism, we can also generate phase 2 from this small resolution. 
 In appendix \ref{app:smallres3} another example of this kind is discussed in more detail. 

%%%%%%%%%%%%%%%%%%%%%%%%%%%%%%%%%%%%
%%%%%%%%%%%%%%%%%%%%%%%%%%%%%%%%%%%%

\section{Flops and the Complete Network of Phases}
\label{sec:Flops}

Both toric and algebraic resolutions only cover part of the phase diagram. In particular, so far we have not realized phases 7 and 8 and their $\Z_2$ counterparts 6 and 5. In this section we will complete the geometric picture by realizing these missing phases in terms of flop transitions along matter curves. 

The interesting point to note is that the flop transitions that will give rise to the missing phases are along matter curves (more accurately, the surfaces that are obtained from the fiber over the matter curves), which are not the intersection of two exceptional divisors\footnote{Note that these flops can of course not be seen from just the 6 phases that appeared in \cite{Esole:2011sm, MS}, which does not include the specific codimension 2 structure of the sections $u_i$ and $v_j$ in (\ref{BinGeo}). }. This observation makes it clear why e.g. phase 8 cannot be realized from a toric resolution. In toric geometry, the resolution is achieved by the triangulation of a fan which defines the toric geometry. Different triangulations can flop a curve which is contained inside two exceptional divisors of an elliptically fibered Calabi-Yau fourfold. Evidently, all the flopped curves among the flop transitions between phase 9, 11, 12 are realized as intersections of Cartan divisors. Contrary to that, the flop transition between phase 9 and phase 8 utilizes a curve which is only inside $D_{-\alpha_2}$. Therefore, that flop cannot be realized in terms of a different triangulation of a fan. In the same spirit, the toric resolution cannot describe flops of a curve which is generated by the decomposition along the $\bar{{\bf 5}}$ matter curve, which is what will be relevant for {the flop transition from phase 8 to 7}.

%%%%%%%%%%%%%%%%%%%%%%%%%%%%%%%%%%%%

\subsection{Contraction Maps and Flops}

Before understanding the flops, we will first of all discuss the contraction of smooth algebraic varieties in detail.
We will construct contraction maps which shrink a smooth algebraic variety $M$ to a point. 
The contraction maps can be constructed by defining surjective holomorphic maps $f_i: X \rightarrow Y$, which are isomorphic maps, except on the locus $M${, which is a subvariety in $X$,} which maps to a point in $Y$\footnote{The contraction maps  are so-called extremal contractions, which map curves that are {on an extremal ray} of the (relative) Mori cone to a point, and are isomorphisms otherwise.}. Here, we assume that $X$ is a smooth algebraic variety. The holomorphic maps are defined patch by patch, so that the maps are consistent on the intersection between the patches. We will choose local holomorphic coordinates in a patch of $Y$ so that the contracted point becomes an origin in the patch {of $Y$}.

Let us see the general strategy for the construction of the contraction maps. First, we pick a patch of $X$, which contains the locus $M$ that will be contracted. Let $f_i$ be holomorphic functions, which vanish on $M$, and we only consider an independent subset of these. 
For example, we will not consider $f_i f_j$ as an element of the set when $f_i$ and $f_j$ are inside the set. Finally, we regard $f_i$ as holomorphic coordinates in a patch of $Y$. Since $f_i$ vanish along the locus $M$, $M$ becomes an origin in this patch of $Y$. Hence, the set of holomorphic functions $f_i$ represent the contraction maps. The (potential) singularity is characterized by constraints among the $f_i$'s. 
For instance, the constraint can be of the type $f_i f_j = f_k f_l$, which is a conifold singularity. In this way we can explicitly construct the flop transitions.

If $f_i$ has a pole at a point $p$, which is not located on $M$, one can generically refine the patches in $X$ as well as the patches in $Y$ so that $f_i$ does not have a pole in the given patch. The point $p$ can be covered by a new patch which does not contain the locus $M$, and we can define a trivial isomorphism in terms of the contraction maps on the new patch, since the contraction maps are one to one maps except for $M$. Therefore, we will neglect poles, which are not on the locus $M$, since they may be generically evaded by the refinement of the patches. However, we have to be careful if there are poles at points on the locus $M$. All the points on the locus $M$ should map to one point in $Y$ by the contraction maps, and one cannot refine the patch in $Y$ to evade the poles. Therefore, we will construct the holomorphic maps $f_i$ which are holomorphic on the {\it whole} locus $M$.

As an example consider the resolved conifold. A resolved conifold can be covered by two patches $U_1$ and $U_2$. We denote the holomorphic coordinates on $U_1$ by $(s_1, s_3, \xi^{\prime})$ and those on $U_2$ by $(s_2, s_4, \xi)$. The transition map between the two patches is 
\be
s_1 = s_2 \xi\,\qquad  s_3 = s_4 \xi\,, \qquad \xi^{\prime} = \frac{1}{\xi} \,.
\ee
Next we construct the contraction of $M= \mathbb{P}^1$ in the resolved conifold. The $\mathbb{P}^1$ is described by $s_1 = s_3 = 0$ in $U_1$ and $s_2 = s_4 = 0$ in $U_2$. The set of holomorphic functions in $U_1$ which vanish on the $\mathbb{P}^1$ is 
\be
\{f_1, f_2, f_3, f_4 \} = \{ s_1, s_3, s_1\xi^{\prime}, s_3\xi^{\prime} \}. \label{contract.conifold}
\ee  
Note that $s_1\xi^{\prime k}$ and $s_3\xi^{\prime k}$ with $k \geq 2$ are not part of this set $f_i$, since it has a pole at $\xi = \frac{1}{\xi^{\prime}} = 0$ such as 
\bea
s_1 \xi^{\prime k} &=& s_2 \xi^{-k+1}, \\
s_3 \xi^{\prime k} &=& s_4 \xi^{-k+1}.
\eea
Hence, \eqref{contract.conifold} is the complete set of the generators of the set of holomorphic functions, which vanish on the $\mathbb{P}^1$. Note that there is a constraint among these functions $f_i$ 
\be 
f_1 f_4 = f_2 f_3 \,,
\ee
which is precisely a  conifold singularity, as expected from the contraction of the $\mathbb{P}^1$. We can also similarly construct the contraction maps in the patch $U_2$.   

We will now apply this general procedure to matter curves in codimension 2. In this way starting with the algebraic resolutions in the last section, we can obtain all phases. The first flop maps phase 11 to 12, which was already realized torically, however not from an algebraic resolution. We can then proceed and realize the remaining phases 7 and 8 by flops starting with phase 9 using the algebraic resolution. In each case we find that the flop is along a matter curve. {We will explicitly construct the new resolved geometry only in one patch. The flop transitions in the other patches will be carried out in a similar manner.}

%%%%%%%%%%%%%%%%%%%%%%%%%%%%%%%%%%%%

\subsection{Flop from Phase 11 to 12 }

Phase 12 is realized by a flop transition starting with phase 11. Recall from section \ref{subsec:Phase11} that phase 11 is realized by 
the small resolution of type $((1,3),(1,1))$.
To pass from phase 11 to 12 the curve that needs to be flopped has weight
\be
{- {\bf w}_3^{\bf 10}= (1, -1,0, 1, -1)}\,,
\ee
which corresponds to {a curve fibered over the matter curve corresponding to $b_1=0$ inside $S$}\footnote{{I.e. this is a so-called matter surface in the fourfold. }} 
\be
\Sigma_{\bf 10}:\qquad \delta_2 = \zeta_1 =b_1=0 \,.
\ee

First consider the patch where we set $\zeta_0 = x = C = y = \delta_5 =1$, i.e. these sections cannot vanish. 
The equation takes the form
\be
b_1+\delta_2 (b_2+\zeta_2)+\zeta_1 (b_3 \delta_2+b_4 \delta_2^2-\zeta_2+b_6 \delta_2^3 \zeta_1) =0 \,.
\ee
Note that the extra condition (\ref{Phase11Extra}) is already used here. 
Again, we can flop the curve by noting that the set of functions vanishing along it is generated by
\be\label{contract.phase11}
\ba
s_1 &= \delta_2\cr
s_2 & = \zeta_1 \cr
s_3& = \delta_2\zeta_2 \cr
s_4 & = \zeta_1\zeta_2 \,.
\ea
\ee
In these coordinates the equation in the current patch can be written as 
\be
b_1 + s_3 + s_1 s_2 b_3 + s_1^2 s_2 b_4 - s_4 + b_2 s_1 + b_6 s_1^3 s_2^2 =0 \,,
\ee
under the condition that
\be
s_1 s_4 = s_2 s_3  \,.\label{conifold.phase11}
\ee
Again, blowing down we arrive at conifold singularities along $s_1 = s_2 = s_3 = s_4 = b_1 = 0$.

Let us check the holomorphicity of the contraction maps defined in \eqref{contract.phase11}. We need to see $\zeta_2 = \infty$, which is possible in the patch where we can set 
\be
x = \zeta_2 = \delta_1 = \delta_2 = \delta_5 = 1.\label{patch.phase11}
\ee
The resolved geometry in this patch is 
\be
y^{\prime 2} + b_1y^{\prime} + b_3y^{\prime}\zeta_0^{\prime 2}\zeta_1^{\prime} = \zeta_1^{\prime} + b_2\zeta_0^{\prime}\zeta_1^{\prime} + b_4\zeta_0^{\prime 3}\zeta_1^{\prime 2} + b_6\zeta_0^{\prime 5}\zeta_1^{\prime 3},
\ee
where we put ${}^{\prime}$ for the variables in the patch \eqref{patch.phase11}. The coordinate transformations between the patches are 
\bea
\zeta_2 &=& \frac{1}{\zeta_0^{\prime}},\\
\zeta_1\delta_2 &=& \zeta_0^{\prime 2} \zeta_1^{\prime},\label{holomorphicity.phase11-2}\\
\left(\zeta_2 + \zeta_1\delta_2(b_6\zeta_1\delta_2 + b_4) + b_2\right)\delta_2 &=& y^{\prime}\label{holomorphicity.phase11-3},
\eea
Therefore, the holomorphicity at $\zeta_2 = \infty$ can be checked by the holomorphicity at $\zeta_0^{\prime} = 0$ in this patch. Note that eq.~\eqref{holomorphicity.phase11-3} can be rewritten as 
\be
\left(\frac{1}{\zeta_0^{\prime}} + b_6(\zeta_0^{\prime 2}\zeta_1^{\prime})^2 + b_2\right)\delta_2 = y^{\prime} - b_4\zeta_0^{\prime 2}\zeta_1^{\prime}.
\ee
The consistency of the equation implies that $\delta_2$ has a zero of order one at $\zeta_{0}^{\prime}$. By combining this result with \eqref{holomorphicity.phase11-2}, $\zeta_1$ also has a zero of order one at $\zeta_0^{\prime}$. Hence, both $\delta_2 \zeta_2$ and $\zeta_1 \zeta_2$ are holomorphic at $\zeta_2 = \infty$.

The conifold singularity \eqref{conifold.phase11} allows us to flop the curve, {which sits in the fiber inside the matter surface, in the usual way of introducing a $\mathbb{P}^1$ with homogeneous coordinates $[\xi_1, \xi_2]$ and resolving by}
\be
s_1 \xi_1  = \xi_2 s_2 \,,\qquad s_3 \xi_1 = \xi_2 s_4    \,.
\ee
In the patch where $\xi_1\not=0$ we consider $\xi = \xi_2/\xi_1$ as well as $s_2$ and $s_4$ as local coordinates, and the flopped geometry is
\be
b_1+b_2\xi s_2+b_3 \xi s_2^2+b_4 \xi^2 s_2^3+b_6 \xi^3 s_2^5+\xi s_4-s_4 =0 \,.
\ee
Next we consider the Cartan divisors. After the resolution, $\zeta_0$ is transformed to $\zeta_0\zeta_1\zeta_2\delta_2 \delta_5$, which in this patch becomes
%except for $\zeta_0=0$ and $\delta_5=0$, which are not in this patch, are
\be 
\zeta_1\zeta_2\delta_2 = s_2s_3 = s_2s_4\xi = 0.
\ee
Hence, we have three Cartan divisors contained in this patch, given by
\bea
D_{-\alpha_0}: &\qquad& s_2 = b_1 + \xi s_4 - s_4 = 0,\\
D_{-\alpha_1}: &\qquad& \xi = b_1 - s_4 = 0,\\
D_{-\alpha_3}: &\qquad& s_4 = b_1 + s_2\xi(b_6s_2^4\xi^2 + b_4s_2^2\xi + b_2 + b_3s_2) = 0 \,.
\eea
Interestingly, $D_{-\alpha_0}$ can be seen in this patch after the flop. Note that we contracted a curve where the infinity point corresponds to $\zeta_0^{\prime} = 0$. Since this point is exactly located on $D_{-\alpha_0}$, we can see a part of $D_{-\alpha_0}$ after the contraction. The Cartan divisor $D_{-\alpha_4}$ can be seen in the other patch where $\xi_2 \neq 0$
\be
D_{-\alpha_4}: \qquad \xi^{\prime} = s_3 + b_2 s_1 + b_1 = 0,
\ee
where $\xi^{\prime} = \frac{\xi_1}{\xi_2}$.

Along the {\bf 10} matter curve, the Cartan divisors $D_{-\alpha_0}$ and $D_{-\alpha_3}$ split
\bea
D_{-\alpha_0}: &\qquad& s_2 = (\xi - 1)s_4 = 0,\\
D_{-\alpha_3}: &\qquad& s_4 = s_2\xi(b_6s_2^4\xi^2 + b_4s_2^2\xi + b_2 + b_3s_2) = 0 \,.
\eea
Note that $s_4 = s_2 = 0$ is the $\P^1$, which appears after the flop, and hence corresponds to the weight ${\bf w}^{\bf 10}_{3} = (1,0,-1,1)$. Moreover, $s_4 = \xi = 0$ corresponds to $-\alpha_1$. Therefore, the splitting along $b_1 = 0$ is 
\be\ba
-\alpha_0 \qquad\longrightarrow\qquad& {\bf w}^{\bf 10}_{3} + (-{\bf w}^{\bf 10}_{10})\cr
-\alpha_3 \qquad\longrightarrow\qquad&  {\bf w}^{\bf 10}_{3} + (-\alpha_1) + (-{\bf w}^{\bf 10}_{5}) \,.
\ea\ee
The splitting along $P=0$ does not change. Hence, the flopped geometry precisely reproduces the splitting structure of phase 12.

\subsection{Flop from Phase 9 to 8}

After this warmup example we now construct the flops to phases that so far had no geometric realization, neither torically nor from the small (algebraic) resolutions. 

Let us first construct the geometric phase corresponding to phase 8. Phase 8 is connected to phase 9 along a codimension 1 wall characterized by 
\be
(0,1,0,-1) \cdot \phi = 0.
\ee
Therefore, we will flop a curve corresponding to the weight ${\bf w}^{\bf 10}_{8} = (0,-1,0,1)$ in phase 9. The expectation of the splitting in phase 8 is that the Cartan divisor $D_{-\alpha_4}$ splits along $b_1 = 0$ as 
\be
-\alpha_4 \quad \rightarrow (-\alpha_1) +  (-\alpha_2) + {\bf w}^{\bf 10}_4 + (- {\bf w}^{\bf 10}_8) \,,
\ee
and the other Cartan divisors are irreducible along $b_1 = 0$. The splitting along $P=0$ is the same as that in phase 9.

Starting with $((1,1),(1,2))$ which is phase 9, and which is summarized in appendix \ref{app:1112} we can flop one of the curves corresponding to the weight ${\bf w}^{\bf 10}_{8}$. 
Let us consider along $b_1=0$ the Cartan divisor $\zeta_2=0$, which is given by
\be
\zeta_2=0:\qquad \delta_1 (\delta_2 + b_3 \zeta_1) =0 \,.
\ee
{Consider the component, which is a matter surface given by}
\be
\Sigma_{{\bf 10}}: \qquad \zeta_2= b_1= (\delta_2 + b_3 \zeta_1) =0 \,.
\ee 
{This matter surface is a fibration over the matter curve $b_1=0$ inside $S$. We will flop the curve that sits in the fiber of this matter surface.}
Consider the patch, which contains this curve, 
\be\label{FlopPatch}
\zeta_0 =x= y = \delta_2 =1 \,.
\ee
{This patch does not contain the Cartan divisors corresponding to} $-\alpha_0$ and $-\alpha_3$. 
The resolved Tate form is
\be\label{TatePatch}
\delta_1 + b_1 + b_3 \zeta_1 \delta_1 = \zeta_2 \zeta_1 (\zeta_2 + b_2 + b_4 \zeta_1 \delta_1 + b_6 \zeta_1^2 \delta_1^2) \,.
%0 = b_1 + \zeta_2 (-\delta_2 \zeta_2 - b_2 - b_4 \delta_1^2) + \delta_1 (\delta_2 + b_3 - b_6 \zeta_2 \delta_1)  \,.
\ee

To contract the curve {in} $\Sigma_{\bf 10}$ we need to write this equation in terms of functions which vanish along it. The ideal of such functions in the patch (\ref{FlopPatch}) is generated by $\zeta_2$, $1 + b_3 \zeta_1$. Furthermore, we explicitly have $\delta_1$ that does not vanish along $\Sigma_{\bf 10}$, so we can multiply each of these functions and obtain another generator $\delta_1 \zeta_2$ and $\delta_1 (1 + b_3 \zeta_1)$. The section $\delta_1$ here may be regarded as a coordinate along the curve which we will contract. In fact, %it is clear from the form (\ref{TatePatch}), that we cannot write the Tate form in terms of this "basis", but 
we need to take $ (1 + b_3 \zeta_1) \rightarrow  (1 + b_3 \zeta_1 - b_6 \zeta_1^3 \zeta_2 \delta_1)$ in order for the holomorphicity at $\delta_1 = \infty$. 
In terms of these functions we define
\be\label{contract1}\ba
s_1 &= \zeta_2 \cr
s_3 &= \delta_1 s_1 =  \delta_1 \zeta_2 \cr
s_2 & = 1 + b_3 \zeta_1 - b_6 \zeta_1^3 \zeta_2 \delta_1\cr
s_4 &= \delta_1 s_2= \delta_1(1 + b_3\zeta_1 - b_6 \zeta_1^3 \zeta_2 \delta_1) \,,
\ea
\ee
so that this is the resolution of 
\be
s_1 s_4= s_2 s_3
\ee
by a $\mathbb{P}^1$ whose affine coordinate is $\delta_1$. 
We can then rewrite the Tate form in this patch by
\be
s_4 + b_1 = \zeta_1(s_1^2 + b_2 s_1 + b_4 \zeta_1 s_3)
%b_1 + s_4 + s_1 (- s_1 (s_2 - b_3 + b_6 s_3) - b_2  ) - b_4 s_3 =0
\ee
with an additional relation
\be
1 + b_3 \zeta_1 -b_6 \zeta_1^3 s_3 = s_2.
\ee
Hence, we have conifold singularities along $s_1 = s_2 = s_3 = s_4 = b_1 = 1 + b_3 \zeta_1 = 0$.

Let us see the holomorphicity of the contraction maps \eqref{contract1} at $\delta_1 = \infty$. In order to see $\delta_1 = \infty$, we will consider a patch where we set 
\be
\zeta_0 = y = \delta_1 = \delta_2 = 1.
\ee
Then, the original Tate form becomes  
\be
1 + b_1 x^{\prime} + b_3 \zeta_1^{\prime} = x^{\prime 3}\zeta_2^{\prime 2}\zeta_1^{\prime} + b_2x^{\prime 2}\zeta_2^{\prime}\zeta_1^{\prime} + b_4x^{\prime}\zeta_2^{\prime}\zeta_1^{\prime 2} + b_6 \zeta_2^{\prime} \zeta_1^{3 \prime}. \label{crTate3}
\ee 
%Eq.~\eqref{crTate3} is already non-singular and we do not need any further resolution. 
We put ${}^\prime$ for the new local coordinates to distinguish them from the original local coordinates. The relations between the previous holomorphic coordinates $(\zeta_1, \zeta_2, \delta_1)$ and $(x^{\prime}, \zeta_1^{\prime}, \zeta_2^{\prime})$ are
\be\label{new.coord1}
\zeta_1 = \zeta_1^{\prime}, \quad \zeta_2 = x^{\prime} \zeta_2^{\prime}, \quad \delta_1 = \frac{1}{x^{\prime}}.
\ee 
Hence, the behavior around $\delta_1 = \infty$ can be understood by looking around $x^{\prime} = 0$ in the new local patch.

By the coordinate transformations \eqref{new.coord1}, $s_1, s_2, s_3, s_4$ in \eqref{contract1} are 
\bea
\zeta_2 &=& x^{\prime}\zeta_2^{\prime},\\
1 + b_3 \zeta_1 - b_6 \zeta_1^3 \zeta_2 \delta_1 &=& 1 + b_3 \zeta_1^{\prime} - b_6 \zeta_1^{\prime 3} \zeta_2^{\prime},\\
\zeta_2 \delta_1 &=& \zeta_2^{\prime},\label{fun1}\\
(1 + b_3\zeta_1 - b_6 \zeta_1^3 \zeta_2 \delta_1) \delta_1 &=& \frac{1 + b_3 \zeta_1^{\prime} - b_6 \zeta_1^{\prime 3}\zeta_2^{\prime}}{x^{\prime}}.\label{fun2}
\eea
%Since the locus $\zeta_1^{\prime} = 0$ in this patch is not on the curve, we will neglect the pole. In other words, that pole can be evaded by the refininement of the patches. 
Eq.~\eqref{fun2} seems to have a pole at $x^{\prime} = 0$ but the numerator $1 + b_3 \zeta_1^{\prime} - b_6 \zeta_1^{\prime 3} \zeta_2^{\prime} $ has a zero at $x^{\prime}$ due to the hypersurface equation \eqref{crTate3}. Therefore both \eqref{fun1} and \eqref{fun2} are holomorphic even at $\delta_1 = \infty$. Note that, because of the subtraction by $b_6 \zeta_1^3 \zeta_2 \delta_1$ in $s_4$, we have the holomorphicity at $\delta_1 = \infty$ in $s_4$. If one multiplies \eqref{fun1} or \eqref{fun2} by another power of $\delta_1$, then those functions have a pole at $x^{\prime} = 0$. %Namely, the holomorphic functions vanishing at $\Sigma_{{\bf w}_8^{10}}$ are generated by \eqref{contract1}. 

The flopped curve is obtained by resolving the conifold singularity by another $\mathbb{P}^1$, which relates 
$s_1\xi_1 = \xi_2 s_2$ and $s_3\xi_1= \xi_2 s_4$, which results in the patch $\xi_1\not=0$ and $\xi= \xi_2/\xi_1$ in
\be\label{P8Patch1}
s_4 + b_1 = \zeta_1 (s_2^2\xi^2 + b_2 s_2\xi + b_4\zeta_1\xi s_4) \,,
%b_6 \xi ^3 s_2^2 s_4-b_3 \xi ^2 s_2^2+b_2 \xi  s_2+b_4 \xi  s_4-b_1+\xi ^2 s_2^3-s_4 =0 \,.
\ee
and the additional relation becomes
\be
1 + b_3\zeta_1 - b_6 \zeta_1^3s_4 \xi = s_2 \,.
\ee
The Cartan divisors can be determined again from considering the locus $\zeta_0=0$ after the resolution, i.e.
\be
\zeta_1 \zeta_2 \delta_1 = \zeta_1 s_3 = \zeta_1 \xi s_4  = 0.
\ee
Hence, we have 
\bea
D_{-\alpha_1}:&\quad& \zeta_1 = s_4 + b_1 = s_2 -1 = 0,\\
D_{-\alpha_2}:&\quad& \xi = s_4 + b_1 = s_2 - 1 - b_3 \zeta_1 = 0,\\
D_{-\alpha_4}:&\quad& s_4 = b_1 - \zeta_1(s_2^2\xi^2 + b_2s_2\xi) = s_2 - 1 - b_3\zeta_1 = 0.
\eea
%Here, the last entry is after the flop. 
Along $b_1=0$ the Cartan divisor $D_{-\alpha_4}$ splits into four components 
$\zeta_1 \xi  s_2 \left(s_2 \xi + b_2 \right) =0$.
One of these is $\zeta_1=0$, which is $D_{-\alpha_1}.[b_1]$, and there is also $\xi = 0$, which is $D_{-\alpha_2}.[b_1]$. The remaining two are matter surfaces with charges that are weights in the ${\bf 10}$ representation. In particular, $s_2 = s_4 = 1+b_3\zeta_1 = 0$ is the $\P^1$ which appears after the flop transition, and hence it corresponds to the weight $-{\bf w}^{\bf 10}_{8} = (0,1,0,-1)$. Therefore, the splitting in this resolution is
\be
-\alpha_4 \quad \longrightarrow \quad -\alpha_1 + (-\alpha_2) + {\bf w}^{\bf 10}_4 + (- {\bf w}^{\bf 10}_8) \,,
\ee
and this exactly recovers the expectation from the gauge theory analysis in phase 8. The splitting of the ${\bf \bar{5}}$ matter curves is completely unaffected and thus the same as in phase 9.

%To get the full picture we also need to consider the patch where we can follow $D_{-\alpha_1}$. For that consider the patch where we can set
%\be
%x=y=\zeta_0 =\delta_2=1 \,.
%\ee
%In fact in this patch it is clear that 
%\be
%D_{-\alpha_4}.[b_1]:\qquad \delta_1 =b_1=0 \quad \Rightarrow  \quad \zeta_1 \zeta_2 (b_2 + \zeta_2)=0\,,
%\ee
%and thus has a $-\alpha_1$ component, as expected. 
%The splitting is this phase, which corresponds to phase 8 is 
%\be
%-\alpha_4 \quad \rightarrow -\alpha_1 -\alpha_2 + {\bf w}^{\bf 10}_4- {\bf w}^{\bf 10}_8 \,.
%\ee
%The splitting of the ${\bf \bar{5}}$ matter curves is completely unaffected and thus the same as in phase 9. 

%%%%%%%%%%%%%%%%%%%%%%%%%%%%%%%%%%%%

\subsection{Flop from Phase 8 to 7}

Finally, we turn to the geometric construction of phase 7. 
 Phase 7 is connected to phase 8 by a codimension 1 wall which is characterized by 
\be
(0,0,-1,1) \cdot \phi = 0.
\ee
The splitting along $b_1=0$ is the same as that in phase 8 but the splitting along $P=0$ changes. In phase 8, the Cartan divisor $D_{-\alpha_3}$ splits, but, in phase 7, the Cartan divisor $D_{-\alpha_4}$ is expected to split as 
\be
{-\alpha_4} \quad \longrightarrow \quad (-{\bf w}^{\bf 5}_4) + {\bf w}^{\bf 5}_5 \,,
\ee
and the other Cartan divisors are irreducible along $P=0$.

Hence, in order to realize phase 7, we need to flop the component of $D_{-\alpha_3}$ in (\ref{Split5bar}) with weight ${\bf w}_{4}^{\bf 5} = (0,0,-1,1)$, in phase 8, which has the equation
\be
P= \delta_2=  \left(b_2 b_3 \delta _1 \zeta _2-b_1 \zeta _2 \left(b_4 \delta _1+b_2 x_2\right)+b_1^2 y\right) =0 \,.
\ee
In this section, we assume that $b_i \neq 0$ which is a generic case even along $P=0$. 

%Note that  since $D_{-\alpha_3}$ remains unchanged in the flop from phase 9 to 8, the same equations determine the curve that needs to be flopped.
For the explicit realization of the flop, we consider the patch in phase 9 where the divisor $D_{-\alpha_3}$ is present, e.g.
\be
\zeta_0 = x = \zeta_1 = \zeta_2 =1  \,.
\ee
Then the equation for the resolved geometry becomes
\be
-b_6 \delta _1^2-b_4 \delta _1+b_3 \delta _1 y+b_1 y-b_2-\delta _2+\delta _1 \delta _2 y^2 =0 \,. \label{phase9-1}
\ee
The curve that has to be flopped has equations
\be
P= \delta_2= b_3 \left(-b_6 \delta _1+y b_3 -b_4\right)+b_1 b_6 =0 \,.
\ee
Note that the divisor $D_{-\alpha_2}$, and hence the curve corresponding to ${\bf w}^{\bf 10}_8$, is not contained in this patch. Therefore, the defining equation of the resolved geometry in phase 8 is the same as that in phase 9.

Let us consider the following holomorphic coordinates that vanish along this curve
\be\label{contraction.map3}
\ba
t_1 &=  \delta_2\cr
t_2 &= b_3 \left(-b_6 \delta _1+y \left(b_3+t_3\right)-b_4\right)+b_1 b_6    \cr
t_3 &= y t_1 \cr
%s_4 &= y s_2  \cr
t_4 &= -b_6 \delta _1^2-b_4 \delta _1+b_3 \delta _1 y+b_1 y-b_2+\delta _1 \delta _2 y^2 \,.
\ea
\ee
$y$ may be regarded as a coordinate along the curve which we will contract. These satisfy a constraint 
\be\label{constraint}\ba
%s_1 s_4 &= s_2 s_3\cr
%0&= -s_1 + s_5 \cr
%-b_6 b_3^2 s_1+b_3^2 s_4-\frac{b_1 b_6 b_3 s_3^2}{s_1}-b_4 b_3 s_2+b_3 s_3 s_4+2 b_1 b_6 s_2-b_2 b_6 b_3^2+b_1 b_4 b_6 b_3-b_1^2 b_6^2-s_2^2=0 \,.
b_3^2b_6t_1 t_4 &= b_3^2 t_2 t_3+b_3 t_3^2 \left(t_2-b_1 b_6\right)-t_1 \left(\left(b_1 b_6-t_2\right) \left(-b_3 b_4+b_1 b_6-t_2\right)+b_2 b_6 b_3^2\right) \,,
\ea\ee
%where the second line is simply the defining equation of the resolved geometry in this patch, and 
which %the relation for $s_1 s_4$ 
follows from the definition of the local coordinates $s_i$. The defining equation of the resolved geometry in this patch becomes
\be\label{defeq8}
-t_1 + t_4 = 0.
\ee

Then, we need to check the holomorphicity at $y = \infty$ of the local coordinates in \eqref{contraction.map3}. $y = \infty$ can be seen in a patch where we set 
\be
\zeta_0 =x= \zeta_1 = y = \xi_2 = 1.
\ee 
The resolved Tate form is 
\be
s_3 \xi^{\prime} + b_1 = s_1^3 \xi^{\prime} - b_3 s_1^2 + b_6 s_1^2 s_3 + b_2 s_1 + b_4 s_3. \label{Tate.phase8-2}
\ee
where $\xi^{\prime} = \xi_1$ after setting $\xi_2 = 1$. The transformations between the coordinates are 
\be
y = \frac{1}{s_1},\quad \delta_1 = \frac{s_3}{s_1}, \quad \delta_2 = s_1^2\xi^{\prime} + b_6s_1s_3 - b_3s_1. 
\ee
Then, $t_1, t_2, t_3, t_4$ become
\bea
t_1 &=& s_1^2\xi^{\prime} + b_6s_1s_3 - b_3s_1,\\
t_2 &=& b_3\xi^{\prime} -b_3b_4 + b_1b_6,\\
t_3&=& s_1\xi^{\prime} + b_6s_3 - b_3,\\
t_4 &=& \frac{-b_4s_3 + b_1 - b_2s_1 + s_3\xi^{\prime}}{s_1}\label{fun3} \,,
\eea
Eq.~\eqref{fun3} seems to have a pole at $s_1 = 0$ at first sight but the numerator has a zero of order two due to the defining equation \eqref{Tate.phase8-2}\footnote{$t_5 = t_4y$ is also a holomorphic function which vanish on the curve with weight $w_1$. However, a new constraint $t_1t_5 = t_3 t_4$ is trivially solved by $t_5 = t_3$ due to the defining equation $-t_1 + t_4 = 0$. Therefore, we do not introduce $t_5$ here.}. Hence, $t_1, t_2, t_3, t_4$ are indeed holomorphic at $y = \infty$.

In fact, combining the equations \eqref{constraint} and \eqref{defeq8}, we can eliminate $t_4$ and arrive at
\be
t_1\left( b_3^2b_6 t_1 + \left(b_1 b_6-t_2\right) \left(-b_3 b_4+b_1 b_6-t_2\right)+b_2 b_6 b_3^2\right)
=t_3 \left( b_3^2 t_2 +b_3 t_3 \left(t_2-b_1 b_6\right)\right) \,,
\ee
which is again a conifold equation $t_1 \sigma_4 = t_3 \sigma_2$.
We can write this, making the locus of the ${\bf 5}$ matter, $P=0$,  more manifest as 
\be\ba
&t_1 \left(b_3^2(b_1b_4b_3-b_2b_3^2)t_1 - {P} \left(-b_3^2t_1- P + 2 b_1 t_2+b_2 b_3^2-b_1 b_4 b_3\right) - b_1 t_2 \left(-b_1 t_2-2 b_2 b_3^2+b_1 b_4 b_3\right)\right) \cr
&= t_3\left(
-b_1 b_3 P t_3
+b_3^2b_1^2t_2 + b_1 b_3 t_3(b_1t_2 + b_2b_3^2 - b_1b_4b_3)\right)
\ea\ee
Therefore, there are conifold singularities along $t_1 = t_2 = t_3 = P = 0$.

The flopped geometry is obtained by considering the small resolution 
\be
t_1 \xi_1 = \xi_2 \sigma_2 \,,\qquad 
t_3 \xi_1 = \xi_2 \sigma_4 \,. 
\ee 
To determine the equation for the Cartan divisors it is necessary to consider the combination (\ref{zeta0trans}) which defines the Cartan divisors\footnote{Recall, that this is the locus that the section $\zeta_0=0$ gets transformed to after the resolution $\zeta_0 \delta_1 \delta_2 \zeta_1\zeta_2=0$, written in this patch, where only $\delta_i$ can be seen.}
\be
\delta_1\delta_2 = t_1t_2-b_3^2t_3-b_3t_3^2+(b_4b_3 - b_1b_6)t_1 = 0.
\ee
Let us consider for instance the patch with $\xi_1\not=0$ and define $\xi = \xi_2/\xi_1$. Then, the flopped geometry in this patch becomes
\be
t_1 = \xi \sigma_2, \qquad t_3 = \xi \sigma_4. \label{phase7-1} \,,
\ee
and the Cartan divisors are characterized by  
\be
\xi(\sigma_2t_2-b_3^2\sigma_4-b_3\xi\sigma_4^2+(b_4b_3 - b_1b_6)\sigma_2) = 0.
\ee
This allows clear identification of the two Cartan divisors: recall that $\delta_2=0$ translates into $t_1=t_3=0$ and thus corresponds to the factor $\xi=0$, the remaining part is the Cartan divisor corresponding to $\delta_1=0$ after the flop
\bea
D_{-\alpha_3}:&\quad & t_1 =t_3 =\xi=0,\label{phase7alpha3}\\
D_{-\alpha_4}:&\quad & (\sigma_2t_2-b_3^2\sigma_4-b_3\xi\sigma_4^2+(b_4b_3 - b_1b_6)\sigma_2) = 0 \quad \hbox{ and } \quad \eqref{phase7-1}\,. \label{phase7alpha4}
\eea
The Cartan divisor $D_{-\alpha_3}$, which originally splits in phase 8, becomes irreducible along $P=0$. On the other hand, the Cartan divisor $D_{-\alpha_4}$ splits along $P=0$. 
To see this, recall that the defining equation of the new component that arises after the flop is given by
\be
S_{f}:\qquad t_1 = t_2 = t_3 = P = 0\,. \label{phase7P1}
\ee
One can see that \eqref{phase7P1} automatically satisfies the restriction of \eqref{phase7alpha4} to $P=0$. This means that $D_{-\alpha_4}.[P]$ is not irreducible but splits with a component \eqref{phase7P1} 
\be
D_{-\alpha_4} \quad \stackrel{P}{\longrightarrow} \quad  S_f +  (D_{-\alpha_4}. [P] - S_f) \,.
\ee
%Furthermore, since we resolved the conifold singularity by inserting the one $\P^1$, $D_{-\alpha_4}.[P]$ may not split into more than two components. 
One can also check that the components \eqref{phase7P1} intersects with the Cartan divisor $D_{-\alpha_3}.[P]$.  
%\ee
%To discuss the entry (*), which is the fate of $D_{-\alpha_4}$, note that $\delta_1=0$ implies
%\be
%\delta_1 =P=0:\qquad b_1t_2 - b_3 (t_3 (b_2+ b_1t_1)+b_1b_3 t_1) =0 \,.
%\ee
%Furthermore, $t_1 = \xi \sigma_2$ and $t_3 = \xi \sigma_4$, which implies that 
%\be
%\delta_1=P=0:\quad 
%\ba
%t_3 = \xi \sigma_4 &= \xi \left(b_3^2(b_1b_4b_3-b_2b_3^2)t_1 - b_1t_2 \left(-b_1 t_2-2 b_2 b_3^2+b_1 b_4 b_3\right)\right)\cr
%t_1= \xi \sigma_2&= b_1 b_3 \xi\left(b_3b_1t_2 + t_3\left(b_1t_2+ b_2b_3^2-b_1b_4b_3\right)\right)
%\ea\ee
%\Hnote{check the following logic again since there was a computation mistake}which implies that $\delta_1=P=0$ has two components, one of which is $\sigma_2=0$. 
%The component $\sigma_2=0$ has intersection with $D_{-\alpha_3}$, whereas the other component does not. 
Therefore, the splitting in this phase is completely consistent with
\be
{-\alpha_4} \quad \rightarrow \quad -{\bf w}^{\bf 5}_4 + {\bf w}^{\bf 5}_5 \,.
\ee
This is precisely what we need for phase 7. Note that the remaining {\bf 10} splittings are as in phase 8 and 9, which is consistent with phase 7.

%%%%%%%%%%%%%%%%%%%%%%%%%%%%%%%%%%%%

%%%%%%%%%%%%%%%%%%%%%%%%%%%%%%%%%%%
%%%%%%%%%%%%%%%%%%%%%%%%%%%%%%%%%%%

\subsection*{Acknowledgements}

\noindent
{We thank Andreas Braun, Thomas Grimm, Kenji Hashimoto, Hee-Cheol Kim, Hans Jockers, Seung-Joo Lee, Joe Marsano, Christoph Mayrhofer and Timo Weigand for discussions, and the Bethe Center in Bonn (where this work was initiated) for hospitality and financial support.}
This work is supported in part by STFC. 
%%%%%%%%%%%%%%%%%%%%%%%%%%%%%%%%%%%
%%%%%%%%%%%%%%%%%%%%%%%%%%%%%%%%%%%

\newpage
%%%%%%%%%%%%%%%%%%%%%%%%%%%%%%%%%%%%
%%%%%%%%%%%%%%%%%%%%%%%%%%%%%%%%%%%%

\appendix
%%%%%%%%%%%%%%%%%%%%%%%%%%%%%%%%%%%%
%%%%%%%%%%%%%%%%%%%%%%%%%%%%%%%%%%%%

\section{Flop Transitions between Toric Resolutions}
\label{app:ToricAlg}

In this appendix, we will algebraically describe the three toric resolutions in section \ref{sec:toric}. The Toric Resolution I can be obtained by the following succession of resolutions \cite{Bershadsky:1996nh}
\be\label{toricI}
\ba
(x, y, z) &\rightarrow& (xe_1, ye_1, e_0e_1),\\
(y, e_1) &\rightarrow& (ye_4, e_1e_4),\\
(x, e_4) &\rightarrow& (xe_2, e_4e_2),\\
(y, e_2) &\rightarrow& (ye_3, e_2e_3),
\ea
\ee
where we use the same characters after the resolutions for notational simplicity. Indeed, one can recover \eqref{resolution} by repeating all the resolution processes \eqref{toricI} as well as the Stanley--Reisner ideal \eqref{tSR} from the projective relations followed from \eqref{toricI}.

\begin{figure}[tb]
\begin{center}
\includegraphics[width=80mm]{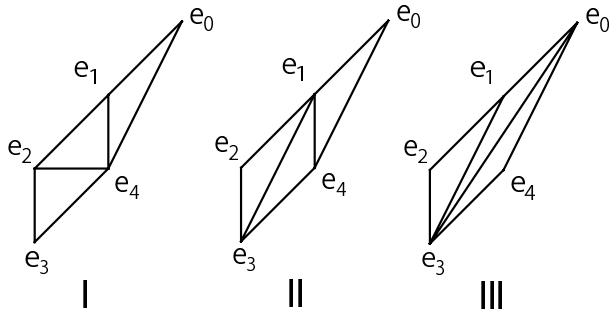}
\end{center}
\caption{The flop transitions among the three toric resolutions. We depict the points $e_0, e_1, e_2, e_3, e_4$ in the two-dimensional space corresponding to the first and the second components of the vectors listed in  \eqref{toric}. }
\label{fig:flop_toric}
\end{figure}

The other two toric resolutions can be obtained by the flop transitions from the Toric Resolution I. Note that the difference of the Stanley--Reisner ideal between the three phases is characterized by the difference of the triangulations of a plane specified the vertices $e_0, e_2, e_3, e_4$. One can see from the triangulations that the transition between the Toric Resolution I and II and the transition between the Toric Resolution II and III are both flop transitions of resolved conifolds as in Figure \ref{fig:flop_toric}.
%%%
%%%%

Let us see this aspect algebraically. In order to see the flop transitions from Toric Resolution I, we consider a polyhedron specified by the vectors in \eqref{toric} with a face spanned by $e_0, e_1, e_2, e_3, e_4$ but now the face is not triangulated. Schematically, we do not draw a line between the points $e_1$ and $e_4$ nor a line between the points $e_2$ and $e_4$ in Toric Resolution I. Then, the toric ambient space has a singularity which is characterized by an algebraic equation
\be 
WY^2 = XZ,
\label{toric.sing}
\ee    
in a patch corresponding to a cone which is dual to a cone spanned by the vectors $e_0, e_2, e_3, e_4$. $W, X, Y, Z$ are local coordinates which are expressed in the original variables as 
\be
W = \frac{e_3e_4y}{xw},\quad X = \frac{e_0^2e_1e_4w^2}{x}, \quad Y = \frac{e_0e_1e_2xw}{y}, \quad Z = \frac{e_1e_2^2e_3x^2}{yw}. \label{local_toric}
\ee
Note that $x, y, w$ do not become zero in this patch. In these local coordinates \eqref{local_toric}, the Tate form \eqref{rTate} becomes
\be
W + b_1 + b_3 X = Z + b_2 Y + b_4 X Y + b_6 X^2 Y. 
\label{def.eq.Tate}
\ee

The resolution of the singularity in \eqref{toric.sing} yields the three toric resolutions in section \ref{sec:toric}. Note that we have an $A_1$ singularity at $X = Y = Z = 0$. There are two difference resolutions for the $A_1$ singularity. First, let us resolve a locus $Y = Z = 0$ 
\be 
Y \eta_1 = Z \eta_2. \label{resolution_toric1}
\ee   
This implies that 
\be
\eta :=  \frac{\eta_1}{\eta_2} = \frac{x e_2 e_3}{w^2 e_0},
\ee
in a patch where $\eta_2 \neq 0$. Hence, neither $e_2$ and $e_0$ nor $e_3$ and $e_0$ vanish simultaneously. By looking into the relation between the Stanely--Reisner ideal and the toric resolutions \eqref{tSR}, the resolution \eqref{resolution_toric1} corresponds to either Toric Resolution I or Toric resolution II.

After the resolution \eqref{resolution_toric1}, the singularity \eqref{toric.sing} becomes
\be 
W Y = X \eta,
\label{conifold1}
\ee
in a patch where $\eta_2 \neq 0$ after the proper transformation. Hence, we have a conifold singularity and there are two ways to resolve it. Let us first resolve the conifold singularity \eqref{conifold1} by %The two resolutions of the conifold singularity \eqref{conifold1} in fact give rise to the Toric Resolution I and II. The Toric Resolution I corresponds to a resolution
\be
W \sigma_1 = X \sigma_2, \quad Y \sigma_2 = \eta \sigma_1. \label{conifold1-1}
\ee
Then, we have 
\be
\frac{\sigma_2}{\sigma_1} = \frac{y e_3}{w^3 e_0^2e_1}.
\ee
Therefore, $e_1$ and $e_3$ do not simultaneously vanish in addition to the Stanley--Reisner ideal $e_0e_2, e_0e_3$ from the resolution \eqref{resolution_toric1}. Hence, the resolutions \eqref{resolution_toric1} and \eqref{conifold1-1} give rise to Toric Resolution I due to the relation \eqref{tSR}. On the other hand, the other resolution of the conifold singularity \eqref{conifold1} becomes
%and the Toric Resolution II corresponds to a resolution 
\be
Y \rho_1 = X \rho_2, \quad W \rho_2 = \eta \rho_1.\label{conifold1-2}
\ee
This implies that 
\be
\frac{\rho_2}{\rho_1} = \frac{x^2 e_2}{yw e_0e_4}. 
\ee
Hence, $e_2$ and $e_4$ do not vanish simultaneously in addition to the Stanley--Reisner ideal $e_0e_2, e_0e_3$ from the resolution \eqref{resolution_toric1}. Therefore, the resolutions \eqref{resolution_toric1} and \eqref{conifold1-2} yield Toric Resolution II due to the relation \eqref{tSR}

Let us then consider the other resolution of the singularity \eqref{toric.sing} along $Y = X = 0$. Namely we consider
\be
Y \rho_1 = X \rho_2. \label{resolution_toric2}
\ee
This implies that 
\be
\rho := \frac{\rho_1}{\rho_2} = \frac{yw e_0e_4}{x^2 e_2}. 
\ee
Therefore, neither $e_0$ and $e_2$ nor $e_2$ and $e_4$ simultaneously vanish. Then, the resolution \eqref{resolution_toric2} yields either Toric Resolution II or Toric Resolution III because of the relation \eqref{tSR}.

After the resolution \eqref{resolution_toric2}, the singularity \eqref{toric.sing} becomes
\be
W Y = Z \rho, \label{conifold2}
\ee
in a patch where we have a local coordinate $\rho = \frac{\rho_1}{\rho_2}$. Again, we encounter a conifold singularity. The conifold singularity \eqref{conifold2} can be resolved in two ways. The first resolution can be done by 
\be
Y \eta_1 = Z \eta_2, \quad W \eta_2 = \rho \eta_1, \label{conifold2-1}
\ee
which implies that 
\be
\eta :=  \frac{\eta_1}{\eta_2} = \frac{x e_2 e_3}{w^2 e_0}.
\ee
Hence, $e_0$ and $e_3$ do not vanish simultaneously in addition to the Stanley--Reisner ideal $e_0e_2, e_2e_4$ from the resolution \eqref{resolution_toric2}. Therefore, the resolutions \eqref{resolution_toric2} and \eqref{conifold2-1} correspond to Toric Resolution II due to the relation \eqref{tSR}. The other resolution of the conifold singularity \eqref{conifold2} can be achieved by 
\be
W \lambda_1 = Z \lambda_2, \quad Y \lambda_2 = \rho \lambda_1. \label{conifold2-2}
\ee
Then, the local coordinate of the inserted $\P^1$ is 
\be
\frac{\lambda_2}{\lambda_1} = \frac{y^2 e_4}{x^3 e_1e_2^2}.
\ee
Hence, $e_1$ and $e_4$ do not vanish simultaneously in addition to the Stanley--Reisner ideal $e_0e_2, e_2e_4$ from the resolution \eqref{resolution_toric2}. Therefore, the resolution \eqref{resolution_toric2} and \eqref{conifold2-2} realizes Toric Resolution III due to the relation \eqref{tSR}.

%%%%%%%%%%%%%%%%%%%%%%%%%%%%%%%%%%%%

%%%%%%%%%%%%%%%%%%%%%%%%%%%%%%%%%%%%
%%%%%%%%%%%%%%%%%%%%%%%%%%%%%%%%%%%%

%%%%%%%%%%%%%%%%%%%%%%%%%%%%%%%%%%%%
%%%%%%%%%%%%%%%%%%%%%%%%%%%%%%%%%%%%

\section{Details of Algebraic Resolutions}
\label{app:AlgRes}

In this appendix, we will summarize how the phases of the algebraic resolutions are determined from the splitting along matter loci in each of the small resolutions $((i, j),(k,l))$. The starting point is the resolution in codimension 1, as determined in section  \ref{subsec:Codim1}. To resolve the elliptic fourfold in higher codimension requires additional small resolutions, as defined in (\ref{SMijkl}), which are labeled by $((i, j),(k,l))$. Depending on which small resolution is chosen, the Cartan divisors of the codimension 1 resolution $D_{-\alpha_i}$ will become reducible along the codimension 2 matter loci, with irreducible components that are curves carrying weights under $SU(5)$ representations. More precisely, the curves have intersections with the Cartan divisors that correspond to weights of the ${\bf 10}$ or ${\bf \bar{5}}$ representations.

In the following we will first  give an argument for the consistency of the types of algebraic resolutions that we consider.
Then we present two examples to demonstrate the algebraic method of determining the phase, and for the remaining small resolutions we will only list the splittings along the codimension 2 loci. 
The final result is summarized in table \ref{AlgResSum}.

%%%%%%%%%%%%%%%%

\subsection{Consistency of resolutions $((i,k)(i,l))$}
\label{subsec:Consistency}

One might naively think that the small resolutions considered here and already in \cite{Lawrie:2012gg}, of the type
\be\label{OurRes}
((i, k), (j, l)) \qquad \hbox{ with }i=j \,,
\ee
which are different from the ones considered in \cite{Esole:2011sm} and \cite{MS}, are not resolutions in the sense, that they are satisfy the condition, that away from singular loci, they are isomorphisms. In this first section we will clarify this point. 

Let us first consider the local geometry around the singularities in \eqref{BinGeo}.  Since all generic singularities after the codimension one resolution appear in the form \eqref{BinGeo}, the analysis of the local patch is enough to see whether a particular type of resolution gives a proper resolution, namely it is an isomorphism away from the singular loci. In this case, we consider $v_1, v_2, u_1, u_2, u_3$ as local coordinates, and they take values at $\C$. There are three lines of conifold singularities
\bea
L_1 &:& v_1 = v_2 = u_1 = u_2 = 0,\\
L_2 &:& v_1 = v_2 = u_1 = u_3 = 0,\\
L_3 &:& v_1 = v_2 = u_2 = u_3 = 0.
\eea 
These three lines intersect at a point $u_1 = u_2 = u_3 = v_1 = v_2 = 0$. 

Consider the resolution of  type $(i,k)(j,l)$ where $i=j$, which are the ones considered here, and in \cite{Lawrie:2012gg}. Since all the coordinates are now on an equal footing, one can consider without loss of generality $(1,1)(1,2)$.
One might now suspect that 
\be\label{RefLoc}
v_1 =v_2= u_2=0 \,,\qquad u_1 u_3 \not=0 \,,
\ee
which is smooth in the geometry before resolution, picks up a full $\mathbb{P}^1$ after the resolution $(1,1)(1,2)$, and thus ceases to be a resolution in the conventional sense. We shall now clarify this point and show that this is in fact not the case, and that these are indeed valid resolutions. 

The first resolution (1,1) is described by 
\be
(v_1, u_1) \rightarrow (v_1^{\prime}\delta_1, u_1^{\prime}\delta_1).
\label{resolve1}
\ee  
The proper transformation of the resolution of \eqref{resolve1} is 
\be
v_1^{\prime}v_2 = u_1^{\prime}u_2 u_3 \label{geometry1}
\ee
We also have a projectivity condition
\be
[v_1^{\prime}, u_1^{\prime}].
\ee
Hence, $v_1^{\prime}$ and $u_1^{\prime}$ are the homogeneous coordinates of the $\P^1$ inserted in the resolution \eqref{resolve1}. The locus $v_1 = u_1 = 0$ where we perform the resolution \eqref{resolve1} is described by $\delta_1=0$ in the new coordinates. However, $\delta_1 = 0$ is not the only equation which defines the $\P^1$. Since $v_1^{\prime}$ and $u_1^{\prime}$ are the homogeneous coordinates of the $\P^1$, they should not be subject to any condition along the locus where the whole $\P^1$ is inserted. Due to the defining equation of the geometry \eqref{geometry1}, the $\P^1$ is inserted along $v_2 = u_2 = 0$ or $v_2 = u_3 = 0$ along the locus $\delta_1 = 0$. These are precisely the loci where the conifold singularities were before the resolution. Namely the $\P^1$ introduced in the resolution \eqref{resolve1} is inserted along $L_1$ or $L_2$.

Note that the geometry \eqref{geometry1} has still a singularity along $v_1^{\prime} = v_2 = u_2 = u_3 = 0$. This corresponds to  a part of $L_3$ which we have not yet resolved. The second step of the resolution of type (1,2) is 
\be
(v_1^{\prime}, u_2) \rightarrow (v_1^{\prime \prime}\delta_2, u_2^{\prime}\delta_2) \label{resolve2}
\ee
The proper transform results in the geometry
\be\label{LocResGeo}
v_1^{\prime\prime} v_2 = u_1^{\prime} u_2^{\prime} u_3
\ee
with projectivity relations
\bea
&&[v_1^{\prime\prime}\delta_2, u_1^{\prime}],\\
&&[v_1^{\prime\prime}, u_2^{\prime}] \,.
\eea
The locus we resolve is characterized by $v_1^{\prime} = u_2 = 0$ before the resolution and it is $\delta_2 = 0$ in the new coordinates. Again, $\delta_2 = 0$ does not fully specify the location where the whole $\P^1$ is. Since the homogeneous coordinates of the $\P^1$ inserted in the second resolution step \eqref{resolve2} are $v_1^{\prime\prime}$ and $u_2^{\prime}$, the locus where the $\P^1$ is inserted should be described by the locus where $v_1^{\prime\prime}$ and $u_2^{\prime}$ are completely free. Hence, in addition to $\delta_2 = 0$, the location sholud be specified by $v_2 = u_3 = 0$. $\delta_2 = v_2 = u_3 = 0$ is exactly the location where we had a singularity after the first resolution. Therefore, the $\P^1$s in the first resolution \eqref{resolve1} and the second resolution \eqref{resolve2} are only inserted along the three singular lines in the original geometry \eqref{BinGeo}.

We can now address the locus of concern \eqref{RefLoc}. 
After the two small resolutions $(1,1)(1,2)$ this locus corresponds to
\be\label{RefLocNew}
\ba
v_1 =  v_1 '' \delta_1 \delta_2 &= 0 \cr
v_2  &= 0 \cr
u_2  = u_2' \delta_2 &=0\cr
u_1 u_3 = u_1' \delta_1 u_3 &\not=0 \,.
\ea\ee
It is clear that this locus does not intersect the $\mathbb{P}^1$ from the first small resolution. 
The one from the second small resolution has projective coordinates $[v_1'',u_2']$. Thus, if at this locus there is a full  $\mathbb{P}^1$  inserted, as one may naively suspect, these coordinates should remain unconstrained. 
To satisfy the above equations while keeping $v_1''$ and $u_2'$ unconstrained, we therefore have
\be
\delta_2= v_2=0 \,.
\ee
However, we also need to satisfy the equation (\ref{LocResGeo}), which implies that the locus (\ref{RefLocNew}) is only part of the geometry if in addition we impose
\be
u_1' u_2' u_3 =0 \quad \stackrel{u_1'  u_3 \not=0}\Rightarrow \quad u_2' =0 \,.
\ee
This shows that along the locus (\ref{RefLocNew}) not a full $\mathbb{P}^1$ is inserted, but only a point on that $\mathbb{P}^1$, which is given by $\delta_2= v_2= u_2'=0$, thus showing that the resolutions of the type (\ref{OurRes}) are completely consistent.

%%%%%%%%%%%%%%%%

\subsection{Small Resolution ((1,1), (1,2)), Phase 4 and  9}
\label{app:1112}

This case was already discussed in detail in \cite{Lawrie:2012gg}, including the intersection computation, so let us just briefly summarize. 
The case $((1,1),(1,2))$ corresponds to the small resolutions 
\begin{equation}
\ba
&(y, \zeta_1; \delta_1) \cr
& (y, \zeta_2; \delta_2)  \,,
\ea
\end{equation}
i.e. we include two new $\mathbb{P}^1$s with corresponding sections $\delta_i$ and homogeneous coordinates $[y, \zeta_i]$. 
The resulting geometry is
\begin{equation}
y \left(\delta _1 \left(b_3 \zeta _1 \zeta _0^2+\delta _2 y\right)+b_1 x\right)=
\zeta _1 \zeta _2 \left(b_2 x^2 \zeta _0+\delta _1 \zeta _1 \zeta _0^3 \left(b_6 \delta _1 \zeta _1 \zeta _0^2+b_4 x\right)+\delta _2 \zeta _2 x^3\right) \,.
\end{equation}
The exceptional divisors are 
\begin{equation}\label{ExDivs}
\begin{array}{c|c}
\hbox{Cartan Divisor} &\hbox{Section}  \cr\hline
D_{-\alpha_0}&  \zeta _0 \\
D_{-\alpha_1}&  \zeta _1  \\
 D_{-\alpha_2}& \zeta _2 \\
 D_{-\alpha_4}& \delta _1 \\
 D_{-\alpha_3}& \delta _2 \end{array}
\end{equation}

To determine the phase that this small resolution realizes, we need to consider the codimension 2 loci. 
Along the {\bf 10} matter locus $b_1=0$ two Cartan divisors split
\begin{equation}
\ba
{-\alpha_2}\quad  &\longrightarrow \quad {-{\bf w}_{6}^{\bf 10}}+ {{\bf w}_{8}^{\bf 10}} \cr
{-\alpha_4} \quad &\longrightarrow \quad  {-{\bf w}_{6}^{\bf 10}}+ {{\bf w}_{4}^{\bf 10} } + (-\alpha_1)\,,
\ea
\end{equation}
where the reducible components labeled by ${\bf w}$ are curves that have intersections with the Cartans given by the weight ${\bf w}$, in the notation of section \ref{sec:Weights}
\begin{equation}\ba
-{\bf w}_{6}^{\bf 10} &= (0, 1, -1, 1, -1) \cr
{\bf w}_{8}^{\bf 10} &= (0, 0, -1, 0, 1) \cr
{\bf w}_{4}^{\bf 10} &= ( 0, 1, 0, 0, -1) \,.
 \ea
\end{equation}
Likewise, along the ${\bf \bar{5}}$ matter locus  $P=b_2 b_3^2+b_1 \left(b_1 b_6-b_3 b_4\right)=0$ the only Cartan divisor that becomes reducible is
\be\label{Split5bar}
{-\alpha_3} \quad \longrightarrow \quad {{\bf w}_4^{\bf 5}} +  (-{\bf w}_3^{\bf 5}) \,,
\ee
where the two weights of the ${\bar{\bf 5}}$ representation are again in the notation of section  \ref{sec:Weights}
\begin{equation}
\ba
{\bf w}_4^{\bf 5}&=(0, 0, 0, -1, 1)\cr
-{\bf w}_3^{\bf 5}&=(0, 0, 1, -1, 0) \,.
\ea
\end{equation}

Combining this information and picking a basis for the relative Mori cone in this small resolution, we can identify it with phase 9 of table \ref{tb:phase}. By reversing the ordering of the Cartan divisors this generates phase 4.

%%%%%%%%%%%%%%%%

\subsection{Small Resolution ((1,1), (2,2)), Phase 3 and 10}
\label{app:1122}

As a second example we consider an example that involves a small resolution including $Y$, i.e. where one of the resolutions are along a "composite" section. The small resolution $((1,1), (2,2))$ is given by
\begin{equation}
\ba
 (y, \zeta_1; \ \delta_1) &\cr
(Y= y+ b_1 x+b_3 \zeta _1 \zeta _0^2, \zeta_2 ;\  \delta_2)& \cr
\ea
\end{equation}
The resolved geometry is
\begin{equation}
y Y = \zeta _1 \zeta _2 \left(b_2 x^2 \zeta _0+\delta _1 \zeta _1 \zeta _0^3 \left(b_6 \delta _1 \zeta _1 \zeta _0^2+b_4 x\right)+\delta _2 \zeta _2 x^3\right)\,.
\end{equation}
Note that unlike in the case $((1,1), (1,2))$ there is an additional constraint among the sections, which has to be taken into account when computing intersections. 
\begin{equation}
b_3 \delta _1 \zeta _1 \zeta _0^2+b_1 x+\delta _1 y =\delta _2 Y \,.
\end{equation}
Taking this into account, the exceptional divisors can be written in terms of two equations 
\begin{equation}
\begin{array}{c|c|c}
\hbox{Divisor} & \hbox{Section} & \hbox{Equation in $Y_4$} \cr\hline
D_{-\alpha_0}& \zeta _0 & 
\begin{array}{c}
 y Y-\delta _2 \zeta _1 \zeta _2^2 x^3 \\
 b_1 x+\delta _1 y-\delta _2 Y
\end{array}
  \cr\hline
D_{-\alpha_1}& \zeta _1 & 
\begin{array}{c}
 Y \\
 b_1 x+\delta _1-\delta _2 Y \ \Rightarrow\  b_1 x+\delta _1
\end{array}
 \cr\hline
D_{-\alpha_3}& \zeta _2 & 
\begin{array}{c}
 y \\
 b_3 \delta _1 \zeta _1+b_1 x-\delta _2+\delta _1 y\ \Rightarrow\   b_3 \delta _1 \zeta _1+b_1 x-\delta _2
\end{array}
 \cr\hline
D_{-\alpha_4}& \delta _1 & 
\begin{array}{c}
 y Y-\zeta _1 \zeta _2 \left(b_2 \zeta _0+\delta _2 \zeta _2\right) \\
 b_1-\delta _2 Y
\end{array}
 \cr\hline
D_{-\alpha_2}& \delta _2 & 
\begin{array}{c}
 y Y-\zeta _1 \zeta _2 \left(b_2 x^2+\delta _1 \zeta _1 \left(b_6 \delta _1 \zeta _1+b_4 x\right)\right) \\
 b_1 x+\delta _1 \left(b_3 \zeta _1+y\right)
\end{array}
\end{array}
\end{equation}

It is clear that along $b_1=0$ the divisors $D_{-\alpha_4}$ and $D_{-\alpha_2}$ will split. 
For example consider $D_{-\alpha_4}$. The defining equations imply immediately that it  splits off one copy of $D_{-\alpha_1}$. The remaining part are weights of the {\bf 10} that follow from the intersection computation. 
In summary we find the following splitting 
\begin{equation}
\ba
-\alpha_2=(0, 1, -2, 1, 0)&  \quad \longrightarrow \quad     (  {0, 1, -1, 1, -1}) +   ({0, 0, -1, 0, 1}) \cr
& \phantom{\quad \longrightarrow \quad} = - {\bf w}_{6}^{\bf 10} +{\bf w}_{8}^{\bf 10} \cr 
-\alpha_4 = (1,0,0,1,-2)  &  \quad \longrightarrow \quad (1, -2, 1, 0, 0) + (0, 1, 0, 0, -1) + (0, 1, -1, 1, -1) \cr
& \phantom{\quad \rightarrow \quad} =  -\alpha_1 +  {\bf w}_{4}^{\bf 10} +(- {\bf w}_{6}^{\bf 10}) \,.
\ea
\end{equation}
Along $P=0$ the following splitting occurs
\begin{equation}\ba
-\alpha_2 =(0, 1, -2, 1, 0) &\quad \longrightarrow \quad (0, 0, -1, 1, 0)+(0, 1, -1, 0, 0) \cr
& \phantom{\quad \longrightarrow \quad} =  {\bf w}_{3}^{\bf 5} +   (-{\bf w}_{2}^{\bf 5})  \,.
\ea
\end{equation}
Comparison with table \ref{tb:phase} yields that this is phase 10 and by reversal of the ordering of the Cartans, phase 3. 

%%%%%%%%%%%%%%%%%

\subsection{Small Resolution $((1,3),(1,2),(1,1))$}
\label{app:smallres3}

In most of this appendix we consider resolutions, where the sections involved in the small resolutions
are $y, Y, \zeta_1$, and $\zeta_2$. One can also consider small resolution with respect to the 
section $C$ in the equation
\begin{equation}
	yY = \zeta_1\zeta_2C \,.
\end{equation}
For example, one can ask about the phase of the space after small resolutions 
$((1,3),(1,1))$. As in section \ref{subsec:Phase11} we calculate the phase in this case by 
applying an additional small resolution $(1,2)$ so as to make all the 
divisors irreducible (or Cartier).  One can ask whether this 
extra small resolution will change the phase. However explicit computation reveals that the additional small resolutions never change the phase. 

In all cases where one performs small resolutions 
with respect to all 3 right hand side coordinates, $\zeta_1, \zeta_2$, and 
$C$, the phases obtained are the same as those occuring in table \ref{AlgResSum}.
Here we include the example of $((1,3),(1,2),(1,1))$ to demonstrate this.
After the 3 small resolutions the geometry is given by the two equations:
\begin{equation}
\ba
		y(y\delta_1\delta_2\delta_3 + b_1x + b_3\zeta_0^2\zeta_1\delta_3) &= \zeta_1\zeta_2C \cr
		x^3\zeta_2\delta_2 + b_2x^2\zeta_0 + b_4x\zeta_0^3\zeta_1\delta_3 + b_6\zeta_0^5\zeta_1^2\delta_3^3& =\delta_1C \,,
		\ea
		\end{equation}
where the $\delta_i$s are introduced sequentially. 
As explained around (\ref{zeta0Extra}), in order to determine the exceptional sections, we need to consider the transformation of $\zeta_0$ under the resolution, which is
\be
\zeta_0 \quad \longrightarrow\quad \zeta_0 \zeta_1 \zeta_2 \delta_2 \delta_3 \,.
\ee
Note that $\delta_1$ is absent, as expected, as this involved the resolution with respect to $C$.
The vanishing locus of the exceptional sections are
\begin{equation}
	\begin{array}{c|c|c}
	\hbox{Section} &\hbox{Root} & \hbox{Equations in resolved geometry}\cr\hline
		\zeta_0 & -\alpha_0&
		\begin{array}{c}
			y(y\delta_1\delta_3 + b_1x) = \zeta_1C \cr
			\delta_1C = x^3 \cr
		\end{array} \cr\hline
		\zeta_1 & -\alpha_1&
		\begin{array}{c}
			\delta_1\delta_2\delta_3 + b_1x = 0 \cr
			\delta_1C = x^3\zeta_2\delta_2 + b_2\zeta_0 \cr
		\end{array} \cr\hline
		\zeta_2 & -\alpha_2&
		\begin{array}{c}
			\delta_1\delta_2 + b_1x + b_3\zeta_1 = 0 \cr
			\delta_1C = b_2x^2 + b_4x\zeta_1 + b_6\zeta_1^2 \cr
		\end{array} \cr\hline
%		\delta_1 & 
%		\begin{array}{c}
%			y(b_1x + b_3\zeta_0^2\zeta_1\delta_3) = \zeta_1\zeta_2C \cr
%			x^3\zeta_2\delta_2 + b_2x^2\zeta_0 + b_4x\zeta_0^3\zeta_1\delta_3 + b_6\zeta_0^5\zeta_1^2\delta_3^2 = 0 \cr
%		\end{array} \cr\hline
		\delta_2 & -\alpha_3&
		\begin{array}{c}
			y(b_1x + b_3\zeta_1\delta_3) = \zeta_1\zeta_2 \cr
			\delta_1 = b_2x^2 + b_4x\zeta_1\delta_3 + b_6\zeta_1^2\delta_3^2 \cr
		\end{array} \cr\hline
		\delta_3 &  -\alpha_4&
		\begin{array}{c}
			b_1y = \zeta_1 \cr
			\delta_1 = \delta_2 + b_2\zeta_0 \cr
		\end{array} 
	\end{array}
\end{equation}

Cartan charges, which are simple roots, associated to these sections (modulo the choice (\ref{Z2Choice})) were computed from the standard intersection computations, see e.g. \cite{MS, Lawrie:2012gg}.
Along the codimension 2 locus of ${\bf 10}$ matter, $b_1 = 0$, some of these divisors become 
reducible, and the Cartan charges split as
\begin{equation}
	\begin{aligned}
		&-\alpha_1 = (1,-2,1,0,0) &\quad\longrightarrow\quad &(0,-1,0,0,1) + (0,-1,1,-1,1) + (1,0,0,1,-2) \cr
		&&& = -{\bf w}^{\bf 10}_4 + {\bf w}^{\bf 10}_6 + (- \alpha_4) \cr
		&-\alpha_3 = (1,-2,1,0,0) &\quad\longrightarrow\quad &(0,-1,1,-1,1) + (0,1,0,-1,0) \cr
		&&& = {\bf w}^{\bf 10}_6 + (- {\bf w}^{\bf 10}_5) \,.\cr
	\end{aligned}
\end{equation}
Along $b_1 = b_3 = 0$ there is a further splitting
\begin{equation}
	\begin{aligned}
		&-\alpha_2 = (0,1,-2,1,0) &\quad\longrightarrow\quad &(0,-1,0,0,1) + 2 \times (0,0,-1,1,0) \cr
		&&& = -{\bf w}^{\bf 10}_5 + 2\times {\bf w}^{\bf 5}_3 \,, \cr
	\end{aligned}
\end{equation}
and along $b_1 = b_2 = 0$ the splitting is
\begin{equation}
	\begin{aligned}
		&-{\bf w}^{\bf 10}_4 = (0,-1,0,0,1) &\quad\longrightarrow\quad &(0,-1,1,-1,1) + (0,0,-1,1,0) \cr
		&&& = {\bf w}^{\bf 10}_6 + {\bf w}^{\bf 5}_3 \,. \cr
	\end{aligned}
\end{equation}
This information, combined with section \ref{sec:Weights}, implies that this resolution, $((1,3),(1,2),(1,1))$ is 
phase $4$, or, with the opposite ordering of the Cartans, phase $9$. It is noteworthy that  the phase of $((1,3),(1,2),(1,1))$ is not the same as the phase for $((1,2),(1,1))$, but related by a flop. This in particular reinforces our observation, that the initial small resolutions determine the phase, and any additional small resolutions amount to relabelings.

%%%%%%%%%%%%%%%%%%%%%

\subsection{Small Resolutions and Phases}

We now summarize the remaining network of small resolutions and associate a gauge theory phase to them. We 
consider only the resolutions that do not involve resolving along the $C = 0$ locus. As stated above, the 
resolutions involving this locus always leave one Cartan divisor reducible. To separate these two reducible 
parts of the divisor one can do another small resolution along that divisor. In the example \ref{app:smallres3} it is 
the $\zeta_1$ divisor which is still reducible after the first two {resolutions}, so we do an additional small 
resolution, $(1,1)$, which makes all Cartan divisors irreducible. The phases arising from performing 3 small 
resolutions are always the same as the phases occurring in table \ref{AlgResSum}.

%%%%%%%%%%%%%%%%%%%%%
\subsubsection*{((1,2), (1,1)), Phase 2 and 11}

Consider the small resolutions 
\begin{equation}
	\begin{aligned}
		&(y, \zeta_2; \delta_1) \cr
		&(y, \zeta_1; \delta_2) \,.\cr
	\end{aligned}
\end{equation}
%From the resulting geometry the exceptional divisors are
%\begin{equation}
%	\begin{array}{l|l|l}
%		\text{Divisor} & \text{Section} & \text{Equation in $Y_4$} \cr\hline
%		D_{-\alpha_0} & \zeta_0 & 0 = y^2\delta_2 + b_1xy - x^3\zeta_1 \cr
%		D_{-\alpha_1} & \zeta_1 & 0 = \delta_1\delta_2 + b_1x \cr
%		D_{-\alpha_2} & \zeta_2 & 0 = \delta_1 + b_1x + b_3\zeta_1 \cr
%		D_{-\alpha_3} & \delta_1 & 0 = b_1xy + b_3y\zeta_0^2\zeta_1\delta_2 
%			- \zeta_1\zeta_2(b_2x^2 + \zeta_1\delta_2(b_4x + b_6\zeta_1\delta_2)) \cr
%		D_{-\alpha_4} & \delta_2 & 0 = b_1y - \zeta_1(\delta_1 + b_2\zeta_0) \cr
%	\end{array}
%\end{equation}
Along the locus of ${\bf 10}$ matter the divisors that split are
\begin{equation}
	\begin{aligned}
		&-\alpha_1 = (1,-2,1,0,0) &\quad\longrightarrow\quad  &(0,-1,1,-1,1) + (1,-1,0,1,-1) \cr
		&&& = {\bf w}^{\bf 10}_6 + (- {\bf w}^{\bf 10}_3) \cr
		&-\alpha_3 = (0,0,1,-2,1) &\quad\longrightarrow\quad  &(0,-1,1,-1,1) + (0,1,0,-1,0) \cr
		&&& = {\bf w}^{\bf 10}_6 + (- {\bf w}^{\bf 10}_5) \cr
		&-\alpha_4 = (1,0,0,1,-2) &\quad\longrightarrow\quad  &(1,-1,0,1,-1) + (0,1,0,0,-1) \cr
		&&& = -{\bf w}^{\bf 10}_3 + {\bf w}^{\bf 10}_4 \,.\cr
	\end{aligned}
\end{equation}
Along the locus of ${\bf 5}$ matter the $D_{-\alpha_3}$ divisor becomes irreducible and 
splits into two divisors with charges
\begin{equation}
	\begin{aligned}
		&-\alpha_3 = (0,0,1,-2,1) &\quad\longrightarrow\quad &(0,0,1,-1,0) + (0,0,0,-1,1) \cr
		&&& = -{\bf w}^{\bf 5}_3 + {\bf w}^{\bf 5}_4 \,.\cr
	\end{aligned}
\end{equation}
This corresponds to phase 11. By reversing the order of the roots we get phase 2.

%%%%%%%%%%

\subsubsection*{((2,2), (2,1)), Phase 2 and 11}

((2, 2), (2, 1)) corresponds to the small resolution
\begin{equation}
	\begin{aligned}
		&(Y, \zeta_2; \delta_1) \cr
		&(Y, \zeta_1; \delta_2) \,.\cr
	\end{aligned}
\end{equation}
Along the ${\bf 5}$ matter locus the Cartan divisor that splits is
\begin{equation}
	\begin{aligned}
		&-\alpha_3 = (0,0,1,-2,1) &\quad\longrightarrow\quad &(0,0,1,-1,0) + (0,0,0,-1,1) \cr
		&&& = -{\bf w}^{\bf 5}_3 + {\bf w}^{\bf 5}_4 \,.\cr
	\end{aligned}
\end{equation}
Along the ${\bf 10}$ locus several Cartan divisors split
\begin{equation}
	\begin{aligned}
		&-\alpha_1 = (1,-2,1,0,0) &\quad\longrightarrow\quad &(0,-1,1,-1,1) + (1,-1,0,1,-1) \cr
		&&& = {\bf w}^{\bf 10}_6 + (- {\bf w}^{\bf 10}_3) \cr
		&-\alpha_3 = (0,0,1,-2,1) &\quad\longrightarrow\quad &(0,-1,1,-1,1) + (0,1,0,-1,0) \cr
		&&& = {\bf w}^{\bf 10}_6 + (- {\bf w}^{\bf 10}_5) \cr
		&-\alpha_4 = (1,0,0,1,-2) &\quad\longrightarrow\quad &(1,-1,0,1,-1) + (0,1,0,0,-1) \cr
		&&& = -{\bf w}^{\bf 10}_3 + {\bf w}^{\bf 10}_4 \,.\cr
	\end{aligned}
\end{equation}
This is again phase 11, and by reversal of the order of the Cartan divisors, we obtain phase 2.

%%%%%%%%%%%%%%%%%%%%%

\subsubsection*{((2,1), (2,2)), Phase 4 and 9}

Interestingly, reversing the order of the small resolutions from $((2,2),(2,1))$ yields a different phase. 
The $((2, 1), (2, 2))$ resolution corresponds to  
\begin{equation}
\ba
&		(Y, \zeta_1; \delta_1) \cr
&		(Y, \zeta_2; \delta_2) \,.
\ea
\ee
%The exceptional divisors after doing these resolutions are
%\begin{equation}
%	\begin{array}{l|l|l}
%		\text{Divisor} & \text{Section} & \text{Equation in $Y_4$} \cr\hline
%		D_{-\alpha_0} & \zeta_0 & 0 = Y^2\delta_1 - b_1xY - x^3\zeta_1 \cr
%		D_{-\alpha_1} & \zeta_1 & 0 = \delta_1 - b_1x \cr
%		D_{-\alpha_2} & \zeta_2 & 0 = \delta_1(\delta_2 - b_3\zeta_1) - b_1x \cr
%		D_{-\alpha_3} & \delta_2 & 0 = b_1xY + b_2x^2\zeta_2 + \delta_1(b_3Y + b_4x\zeta_2 + b_6\zeta_2\delta_1) \cr
%		D_{-\alpha_3} & \delta_1 & 0 = b_1Y + \zeta_1\zeta_2(b_2\zeta_0 + \zeta_2\delta_2) \cr
%	\end{array}
%\end{equation}
Along the ${\bf 10}$ the splitting is
\begin{equation}
	\begin{aligned}
		&-\alpha_2 = (0,1,-2,1,0) &\quad\longrightarrow\quad &(0, 1,-1, 1, -1) + (0,0,-1,0,1) \cr
		&&& = -{\bf w}^{\bf 10}_6 + {\bf w}^{\bf 10}_8 \cr
		&-\alpha_4 = (1,0,0,1,-2) &\quad\longrightarrow\quad &(1,-2,1,0,0) + (0,1,-1,1,-1) + (0,1,0,0,-1) \,.\cr
		&&& = -\alpha_1 + (-{\bf w}^{\bf 10}_6) + {\bf w}^{\bf 10}_4 \,. \cr
	\end{aligned}
\end{equation}
At the ${\bf 5}$ matter locus the $D_{-\alpha_3}$ divisor splits
\begin{equation}
	\begin{aligned}
		&-\alpha_3 = (0,0,1,-2,1) &\quad\longrightarrow\quad &(0,0,1,-1,0) + (0,0,0,-1,1) \cr
		&&& = -{\bf w}^{\bf 5}_3 + {\bf w}^{\bf 5}_4 \,.\cr
	\end{aligned}
\end{equation}
This corresponds to phase 9, and by reversal of the Cartan divisor ordering, phase 4.

%%%%%%%%%%%%%%%%%%%%%

\subsubsection*{((1,1), (1,2)), Phase 4 and  9}

This small resolution was discussed earlier in appendix \ref{app:1112}.

%%%%%%%%%%%%%%%%%%%%%

\subsubsection*{((1,1), (2,2)), Phase 3 and 10}

This small resolution was discussed earlier in appendix \ref{app:1122}.

%%%%%%%%%%%%%%%%%%%%%

\subsubsection*{((2,1), (1,2)), Phase 3 and 10}
The small resolutions are
\begin{equation}
	\begin{aligned}
		&(Y, \zeta_1; \delta_1) \cr
		&(y, \zeta_2; \delta_2) \,.\cr
	\end{aligned}
\end{equation}
Along the ${\bf 10}$ matter curve the splitting of the charges is
\begin{equation}
	\begin{aligned}
		&-\alpha_2 = (0,1,-2,1,0)  &\quad\longrightarrow\quad &(0,1,-1,1,-1) + (0,0,-1,0,1) \cr
		&&& = -{\bf w}^{\bf 10}_6 + {\bf w}^{\bf 10}_8 \cr
		&-\alpha_4 = (1,0,0,1,-2)  &\quad\longrightarrow\quad &(0,1,-1,1,-1) + (0,1,0,0,-1) + (1,-2,1,0,0) \cr
		&&& = -{\bf w}^{\bf 10}_6 + {\bf w}^{\bf 10}_4 + (- \alpha_1) \,.\cr
	\end{aligned}
\end{equation}
Along the ${\bf 5}$ matter curve the splitting is
\begin{equation}
	\begin{aligned}
		&-\alpha_2 = (0,1,-2,1,0) &\quad\longrightarrow\quad &(0,0,-1,1,0) + (0,1,-1,0,0) \cr
		&&& = {\bf w}^{\bf 5}_3  + (- {\bf w}^{\bf 5}_2) \,. \cr
	\end{aligned}
\end{equation}

%%%%%%%%%%%%%%%%

\subsubsection*{((1,2), (2,1)), Phase 3 and 10}

This resolution was done in \cite{MS}. 
The splittings were shown to be as follows: \\
Along $b_1=0$ the divisors $D_{-\alpha_4}$ and $D_{-\alpha_2}$ will split:
\begin{equation}
	\begin{aligned}
		&-\alpha_2 = (0,1,-2,1,0)  &\quad\longrightarrow\quad &(0,1,-1,1,-1) + (0,0,-1,0,1) \cr
		&&& = -{\bf w}^{\bf 10}_6 + {\bf w}^{\bf 10}_8 \cr
		&-\alpha_4 = (1,0,0,1,-2)  &\quad\longrightarrow\quad &(0,1,-1,1,-1) + (0,1,0,0,-1) + (1,-2,1,0,0) \cr
		&&& = -{\bf w}^{\bf 10}_6 + {\bf w}^{\bf 10}_4 + (- \alpha_1) \,.\cr
	\end{aligned}
\end{equation}
Along $P=0$ the splitting is
\begin{equation}
	\begin{aligned}
		&-\alpha_2 = (0,1,-2,1,0) &\quad\longrightarrow\quad &(0,0,-1,1,0) + (0,1,-1,0,0) \cr
		&&& = {\bf w}^{\bf 5}_3 + (- {\bf w}^{\bf 5}_2) \,. \cr
	\end{aligned}
\end{equation}

\subsubsection*{((2,2), (1,1)), Phase 3 and 10}
The resolutions are
\begin{equation}
	\begin{aligned}
		&(Y, \zeta_2; \delta_1) \cr
		&(y, \zeta_1; \delta_2) \,.\cr
	\end{aligned}
\end{equation}
Along $b_1=0$:
\begin{equation}
	\begin{aligned}
		&-\alpha_2 = (0,1,-2,1,0) &\quad\longrightarrow\quad &(0,1,-1,1,-1) + (0,0,-1,0,1) \cr
		&&& = -{\bf w}^{\bf 10}_6 + {\bf w}^{\bf 10}_8 \cr
		&-\alpha_4 = (1,0,0,1,-2) &\quad\longrightarrow\quad &(0,1,-1,1,-1) + (1,-1,1,0,-1) \cr
		&&& = -{\bf w}^{\bf 10}_6 + {\bf w}^{\bf 10}_7 \,.\cr
	\end{aligned}
\end{equation}
Along $P=0$:
\begin{equation}
	\begin{aligned}
		&-\alpha_2 = (0,1,-2,1,0) &\quad\longrightarrow\quad &(0,0,-1,1,0) + (0,1,-1,0,0) \cr
		&&& = {\bf w}^{\bf 5}_3 + (- {\bf w}^{\bf 5}_2) \,. \cr
	\end{aligned}
\end{equation}

%%%%%%%%%%%%%%%%%%%%%%%%%%%%%%%%%%%%
%%%%%%%%%%%%%%%%%%%%%%%%%%%%%%%%%%%%

%%%%%%%%%%%%%%%%%%%%%%%%%%%%%%%%%%%%
%%%%%%%%%%%%%%%%%%%%%%%%%%%%%%%%%%%%

\newpage

%%%%%%%%%%%%%%%%%%%%%%%%%%%%%%%%%%%%
%%%%%%%%%%%%%%%%%%%%%%%%%%%%%%%%%%%%

%\bibliography{FGUTbib}
%\bibliographystyle{JHEP}

\providecommand{\href}[2]{#2}\begingroup\raggedright\endgroup

%%%%%%%%%%%%%%%%%

\end{document}